\documentclass{JHEP3}
\usepackage{amsmath,amssymb}
\usepackage[dvips]{graphicx}
\newcommand{\beq}{\begin{eqnarray}}
\newcommand{\eeq}{\end{eqnarray}}
\newcommand{\SU}{\text{SU}}
\newcommand{\SO}{\text{SO}}
\newcommand{\Sp}{\text{Sp}}

\newcommand{\U}{\text{U}}
\newcommand{\tr}{\mathop{\mathrm{tr}}}

\title{Universality of Phases in QCD and QCD-like Theories}
\author{Masanori Hanada$^a$ and Naoki Yamamoto$^b$\\
  $^a\,$Department of Physics, University of Washington, 
 Seattle, WA 98195-1560, USA\\
  \mbox{$^b\,$Institute for Nuclear Theory, University of Washington, 
  Seattle, Washington 98195-1550, USA}\\
  Email: \email{mhanada@u.washington.edu}, 
  \email{nyama@u.washington.edu}}

\abstract{We argue that the whole or the part of the phase diagrams of 
QCD and QCD-like theories should be universal in the large-$N_c$ limit 
through the orbifold equivalence. The whole phase diagrams,
including the chiral phase transitions and the BEC-BCS crossover regions, 
are identical between $\SU(N_c)$ QCD at finite isospin chemical potential
and $\SO(2N_c)$ and $\Sp(2N_c)$ gauge theories at finite baryon chemical potential.
Outside the BEC-BCS crossover regions in these theories,
the phase diagrams are also identical to that of
$\SU(N_c)$ QCD at finite baryon chemical potential. 
We give examples of the universality in some solvable cases: 
(i) QCD and QCD-like theories at asymptotically high density 
where the controlled weak-coupling calculations are possible,
(ii) chiral random matrix theories of different universality classes,
which are solvable large-$N$ (large volume) matrix models of QCD. 
Our results strongly suggest that the chiral phase transition and 
the QCD critical point at finite baryon chemical potential can be studied 
using sign-free theories, such as QCD at finite isospin chemical potential, 
in lattice simulations.}


\preprint{INT-PUB-11-012; \today}

\begin{document}
\section{Introduction}
Unraveling the properties of quantum chromodynamics (QCD) at finite temperature $T$ 
and finite baryon chemical potential $\mu_B$ is essential to understand various phenomena
from ultrarelativistic heavy-ion collisions, the early Universe, 
supernova explosion, and neutron stars, to possible quark stars
 (for a recent review, see, e.g., \cite{Fukushima:2010bq}).
A lot of progress have been made theoretically by the first-principles lattice
simulations in QCD at high $T$ and sufficiently small $\mu_B$ 
in connection with the experimental investigation of the quark-gluon plasma in the
ultrarelativistic heavy-ion collisions at RHIC and LHC. On the other hand,
the lattice technique is not available at finite $\mu_B$ because 
of the {\it sign problem}: the fermion determinant 
is no longer real at finite $\mu_B$, and the Monte Carlo approach based on the importance 
sampling does not work. This is why QCD at finite $\mu_B$ has not been fully understood,
such as the precise location of the QCD critical point(s) 
(for a review, see \cite{Stephanov:2004wx})
and the realization of the color superconductivity at intermediate $\mu_B$ 
(for a review, see \cite{Alford:2007xm})
relevant to the physics of neutron stars, etc.

Still there are a class of QCD-like theories which are free from the sign problem.
These theories intensively studied so far include two-color QCD 
with even numbers of fundamental flavors \cite{Kogut:1999iv, Kogut:2000ek}, 
any-color $\SU(N_c)$ QCD with adjoint fermions \cite{Kogut:2000ek},
and $\SU(N_c)$ QCD at finite isospin chemical potential $\mu_I$ \cite{Alford:1998sd, Son:2000xc}.
In addition to the chiral phase transition,
these theories exhibit the Bardeen-Cooper-Schrieffer (BCS) pairing of quarks 
at large chemical potential due to the same mechanism as 
the color superconductivity in real QCD \cite{Alford:2007xm}.
However, it is not clear, at the quantitative level, or even at the qualitative level,
how the phase diagrams of these theories are related to each other, 
and more importantly, to that of real QCD at finite $\mu_B$.
There are also other theories which are free from sign problem at finite $\mu_B$
but have not been well studied:
$\SO(2N_c)$ and $\Sp(2N_c)$ gauge theories with $N_f$ fundamental Dirac fermions.\footnote{The 
symplectic group is defined as $\Sp(2N_c) = \{g \in \U(2N_c) |g^T J_c g = J_c \}$, 
where $J_c$ is the antisymmetric matrix defined in (\ref{eq:J_c}). 
The dimension of $\Sp(2N_c)$ is $N_c(2N_c+1)$.}
Apparently, there is no a priori reason why these theories
capture the physics of $\SU(N_c)$ QCD at finite chemical potential either,
since the structures of gauge groups are different.

\begin{figure}[t]
\begin{center}
\includegraphics[width=11cm]{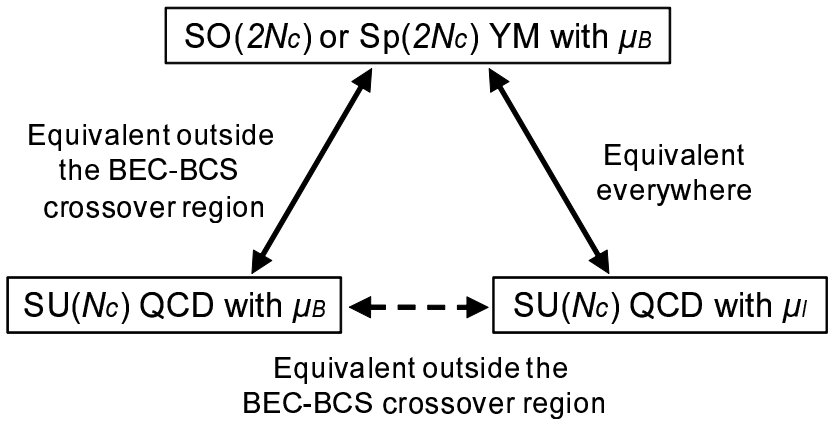}
\end{center}
\vspace{-0.5cm}
\caption{Relations between $\SU(N_c)$ QCD at finite $\mu_B$ ($\mu_I$) and 
$\SO(2N_c)$ and $\Sp(2N_c)$ Yang-Mills (YM) theories at finite $\mu_B$. 
$\SU(N_c)$ QCD at finite $\mu_I$ can be obtained from
$\SO(2N_c)$ or $\Sp(2N_c)$ gauge theory at finite $\mu_B$ 
through the orbifold projection in the whole phase diagram, 
while QCD at finite $\mu_B$ can be obtained outside the BEC-BCS crossover region
of these theories. As a result, QCD at finite $\mu_B$ is equivalent to 
QCD at finite $\mu_I$ outside the BEC-BCS crossover region of the latter.}
\label{fig:QCD}
\end{figure}

In this paper, we argue that the whole or the part of the phase diagrams of 
QCD and QCD-like theories should be {\it universal} in the large-$N_c$ limit via 
the orbifold equivalence.\footnote{The universality of the Wilson loops 
between $\SU(N_c)$, $\SO(2N_c)$, and $\Sp(2N_c)$ in pure gauge theories 
in the large-$N_c$ limit was pointed out long time ago in \cite{Lovelace:1982hz}, 
even before the notion of the orbifold equivalence was found. 
Probably the first paper which considered an equivalence of the phase diagrams
of QCD-like theories is \cite{Unsal:2007fb}, in which 
$\SU(N_c)$ gauge theory with two index fermion representations [adjoint, (anti)symmetric, and bifundamental] 
on the compact space $S^1 \times S^3$ as a function of volume was studied. 
Our paper is the first, to our knowledge, to argue the large-$N_c$ universality 
of the phase diagrams of QCD and sign-free $\SO(2N_c)$ and $\Sp(2N_c)$ gauge theories 
with fundamental fermions at finite chemical potential
on physically relevant space ${\mathbb R}^4$.}
The relations between QCD and QCD-like theories are summarized in Fig.~\ref{fig:QCD}.
The whole phase diagrams described by the chiral condensate and the 
superfluid gap should be quantitatively identical between $\SU(N_c)$ 
QCD at finite $\mu_I$ and $\SO(2N_c)$ and $\Sp(2N_c)$ gauge theories at finite $\mu_B$.\footnote{We
note that the deconfinement phase transition characterized by the Polyakov loop 
should also be identical between these theories. Because the quark chemical potential $\mu$ 
does not affect the gauge dynamics as long as $\mu = {\cal O}(N_c^0)$ in the large-$N_c$ counting, 
the deconfinement temperature is independent of $\mu$ at the leading order; 
the universality at nonzero $\mu$ follows once that at $\mu=0$ is provided \cite{Lovelace:1982hz}. 
For this reason, except Sec.~\ref{sec:projection} where we discuss the $1/N_c$ correction 
to the deconfinement temperature, we shall mostly concentrate on 
the phases related to the chiral and flavor dynamics which presumably depend on $\mu$ 
at the leading order of $1/N_c$ expansion.}
As common properties of these theories, 
the Bose-Einstein condensate (BEC) of the Nambu-Goldstone (NG) modes at small chemical potential,
as well as the BCS-type pairing at large chemical potential, appear in the phase diagrams.
Since the BEC and BCS pairings carry the same quantum numbers and break the same symmetry, 
the BEC and BCS regions should be continuously connected without any phase transition.
This is the BEC-BCS crossover similar to that in nonrelativistic 
condensed-matter systems \cite{Eagles1969, Leggett1980, Nozieres1985}
and the continuity between the hadron phase and the color 
superconducting phase (hadron-quark continuity) in three-flavor QCD at finite $\mu_B$ 
\cite{Schafer:1998ef, Hatsuda:2006ps, Yamamoto:2007ah}.\footnote{The BEC regions 
in these theories appear in a model-independent way 
(see the discussion in Sec.~\ref{sec:class}).
This is in contrast to QCD at finite $\mu_B$ where the BEC of diquark pairing
can appear depending on the details of the models \cite{Kitazawa:2007zs, Abuki:2010jq}.
Throughout this paper, we assume the crossover between the BEC and BCS regions 
similarly to \cite{Son:2000xc, Splittorff:2000mm} though our result of the 
universality of the phase diagrams does not rely on this assumption.}
The universality of the phase diagrams means that 
both the chiral phase transitions and the BEC-BCS crossover regions
should appear at the same coordinates in the $T$-$\mu$ 
plane independently of the theories in spite of the different symmetry breaking patterns.

We also argue that the phase diagrams of these theories outside the BEC-BCS crossover regions
should be identical to that of $\SU(N_c)$ QCD at finite $\mu_B$, which is most relevant in reality.
In particular, the magnitude and temperature dependence of the chiral condensate in QCD 
at finite $\mu_B$ should be exactly the same as those in QCD at finite $\mu_I$.
Since the latter can be obtained by dropping the complex phase of the fermion determinant 
in the former for even number of flavors, this suggests that the phase-quenched approximation 
for the chiral condensate is exact in this region.
Actually, this phenomenon has already been observed in the lattice QCD simulations \cite{Allton:2002zi} 
and the model calculations, such as the chiral random matrix model \cite{Klein:2003fy}, 
the Nambu--Jona-Lasinio model \cite{Toublan:2003tt, Barducci:2005ut},
and the hadron resonance gas model \cite{Toublan:2004ks},
though the reason has been unclear so far.
We note that our arguments based on the orbifold equivalence 
are model-independent (though it is exact in the large-$N_c$ limit), and thus, 
would provide a solid theoretical basis of the phase-quenched approximation.

The idea of the orbifold equivalence first originates from the string theory 
\cite{Kachru:1998ys,Lawrence:1998ja,Bershadsky:1998mb},
and then has been generalized to the quantum field theory without reference to the string theory
\cite{Bershadsky:1998cb,Schmaltz:1998bg,Erlich:1998gb,Strassler:2001fs,
Kovtun:2003hr,Kovtun:2004bz,Kovtun:2005kh,Unsal:2006pj}.
The procedure of orbifolding (or orbifold projection) is as follows.
Identify a discrete global symmetry of the original theory (the ``parent" theory).
Eliminate all the degrees of freedom in the parent theory 
which are not invariant under the discrete symmetry.
This gives a new theory (the ``daughter" theory).
Then a class of correlation functions and observables are shown to be identical between 
the parent and daughter theories in the large-$N_c$ limit,
as long as the symmetry used to make the projection is not broken \cite{Kovtun:2004bz}.
For example, as recently proposed \cite{Cherman:2010jj, Cherman:2011mh}, 
$\SU(N_c)$ QCD with $N_f$ quarks at finite $\mu_B$ can be obtained 
from the $\SO(2N_c)$ gauge theory with $N_f$ fundamental fermions at finite $\mu_B$ 
through an orbifold projection.
Here it is this condition, ``outside the BEC-BCS region" mentioned above, that
$\U(1)$ baryon number symmetry of the parent $\SO(2N_c)$ and $\Sp(2N_c)$ gauge theories, 
which is used for the projection, is not broken spontaneously.

To be precise, the orbifold equivalence has been proven to all orders in the perturbation theory
but not nonperturbatively, except certain QCD-like theories containing adjoint 
scalar or fermion matter at $\mu_B=\mu_I=0$ \cite{Kovtun:2003hr} 
and supersymmetric QCD at finite $\mu_B$ or $\mu_I$ in a holographic setup \cite{Hanada:2012nj}.
In this paper, we provide new evidence for the nonperturbative orbifold equivalence
within real QCD and QCD-like theories with fundamental fermions. 
At sufficiently large chemical potential,
nonperturbative observables, such as the superfluid gap and the diquark condensate,
can be computed using the controlled weak-coupling calculations owing to the asymptotic freedom.
Then we can explicitly compute the $1/N_c$ corrections 
and demonstrate that the equivalence is not only exact in the large-$N_c$ limit
but is well satisfied even for $N_c=3$.
We also apply the idea of the orbifold equivalence to the chiral random matrix theory,
a solvable large-$N$ (large volume) matrix model of QCD 
(for reviews, see \cite{Verbaarschot:2000dy, Akemann:2007rf}).
We verify the nonperturbative orbifold equivalence between 
the random matrix theories of different universality classes
by computing the effective potentials in the $N \rightarrow \infty$ limit (thermodynamic limit).
Our calculations and arguments, though do not constitute a complete proof, 
provide overwhelming evidence for the nonperturbative equivalence in real QCD 
at finite chemical potential.

The paper is organized as follows. In Sec.~\ref{sec:class}, we study the properties of
$\SU(N_c)$, $\SO(2N_c)$, and $\Sp(2N_c)$ gauge theories at finite chemical potential.
In Sec.~\ref{sec:QCD}, after a brief review of the orbifold equivalence 
in the large-$N_c$ QCD, we construct the orbifold projections between QCD and QCD-like theories. 
We then argue that the phase diagrams of these theories should be universal.
Also we compute the $1/N_c$ corrections at asymptotically high density. 
In Sec.~\ref{sec:RMT}, 
we construct the orbifold projections between the chiral random matrix 
theories of the different universality classes.
We then explicitly check the nonperturbative orbifold equivalence.
Section~\ref{sec:conclusion} is devoted to conclusion and discussion.

\section{Classification of QCD and QCD-like theories}
\label{sec:class}
In sections~\ref{sec:class} and \ref{sec:QCD},
we consider the following classes of theories:
$\SU(N_c \geq 3)$, $\SO(2N_c)$, and $\Sp(2N_c)$ Yang-Mills theories with
$N_f$ fundamental Dirac fermions.
For $\SU(N_c)$ gauge theory, we consider finite baryon chemical potential $\mu_B$
or finite isospin chemical potential $\mu_I$. For $\SO(2N_c)$ and $\Sp(2N_c)$
gauge theories, we consider finite baryon chemical potential $\mu_B$.
We can classify these theories by the Dyson index $\beta$ of the Dirac operator
which reflects the number of independent degrees of freedom per matrix element
in corresponding chiral random matrix theories, 
as we will briefly review in Sec.~\ref{sec:RMT}:
$\SU(N_c)$, $\SO(2N_c)$, and $\Sp(2N_c)$ gauge theories correspond to 
$\beta=2$, $\beta=4$, and $\beta=1$, respectively.

The Lagrangian of the gauge theories in the Euclidean spacetime is given by
\begin{eqnarray} 
\label{eq:Lagrangian}
\mathcal{L}_{G} 
=\frac{1}{4 g_{G}^{2} } \tr (F^G_{\mu \nu})^2
+ 
\sum_{f=1}^{N_{f}}
\bar{\psi}^G_{f} ({\cal D} + m) \psi^G_{f},
\end{eqnarray}
where ${G}$ denotes the gauge group $\SU(N_c)$, $\SO(2N_c)$, or $\Sp(2N_c)$
and $f$ denotes the flavor index.
$F^G_{\mu \nu}$ is the field strength of each gauge field $A^{G}_{\mu} = A^G_{\mu a} T^{G}_{a}$ 
with $T^{G}_{a}$ being the generators of each gauge group normalized such that 
$\tr(T^{G}_{a} T^{G}_{b}) = (1/2) \delta_{ab}$.
The Dirac fermion $\psi^G_{f}$ belongs to the fundamental representation of the gauge group $G$,\footnote{For 
$G=\SO(2N_c)$, it is the $2N_c$-component vector representation.}
and $m_f=m$ is the degenerate quark mass.\footnote{The degenerate quark mass 
is not essential in our argument. In order to avoid the sign problem, pairwise $m_f$ is 
necessary in $\SU(N_c)$ QCD at finite $\mu_I$ and $\Sp(2N_c)$ gauge theory at finite $\mu_B$,
but is not in $\SO(2N_c)$ gauge theory at finite $\mu_B$. See the subsections below.} 
The Dirac operator ${\cal D}$ is defined as
\beq
\label{eq:D_mu_B}
{\cal D}=\gamma^{\mu} D_{\mu} + \mu \gamma^{4},
\eeq
for finite quark chemical potential $\mu$, and 
\beq
\label{eq:D_mu_I}
{\cal D}=\gamma^{\mu} D_{\mu} + \frac{1}{2}\mu_I \gamma^{4} \tau^3,
\eeq
for finite isospin chemical potential $\mu_I = 2 \mu$ when $N_f=2$.\footnote{In this paper, 
we consider the isospin chemical potential $\mu_I$ for $N_f=2$ unless otherwise
stated explicitly. 
Our argument can be generalized to any even $N_f$ if we define $\mu_I$ by regarding 
$N_f/2$ quarks as ``up" with the quark chemical potential $\mu$ and $N_f/2$ quarks 
as ``down" with the quark chemical potential $-\mu$.}
Here $D_{\mu} = \partial_{\mu}+i A^{G}_{\mu}$ is the color covariant derivative.
The $\SU(2)$ isospin generators $\tau_i$ are normalized such that
 $\tr({\tau_i \tau_j}) = 2\delta_{ij}$.
In both cases above, the Dirac operator preserves the chiral symmetry $\{\gamma_5,{\cal D}\}=0$.

In this section, we will investigate the properties (the phase diagram in particular) 
of each gauge theory.
For completeness, we repeat some of arguments on $\SU(N_c)$ QCD at finite $\mu_B$ 
\cite{Alford:2007xm} and $\mu_I$ \cite{Son:2000xc} in the literature.
The arguments in this section are independent of $N_c$, except that we consider
$N_c \geq 3$ for $\SU(N_c)$ QCD.

\subsection{QCD: $\SU(N_c \geq 3)$ gauge theory at finite $\mu_B$ ($\beta=2$)}
Let us first consider the $\SU(N_c \geq 3)$ gauge theory with fundamental fermions (QCD).
In the QCD vacuum ($\mu=0$), the Dirac operator ${\cal D}$ is anti-Hermitian
and its eigenvalue $i\lambda_n$ defined by ${\cal D}\psi_n=i\lambda_n \psi_n$
is always pure imaginary, $\lambda_n \in \mathbb{R}$.
From the chiral symmetry $\{\gamma_5,{\cal D}\}=0$, $-i\lambda_n$
is also the eigenvalue when $\lambda_n \neq 0$ and the fermion determinant
$\det({\cal D}+m)$ is always real and nonnegative. 
This allows us to use the standard Monte-Carlo simulation 
based on the importance sampling.
However, this is no longer true in QCD at finite $\mu_B$. 
Because $\lambda_n$ is generally complex and so is the fermion determinant, 
the standard Monte-Carlo simulation technique fails (the fermion sign problem). 
In this case, the Dirac operator is written in the form of the complex matrix 
[see (\ref{eq:Dyson2})], and the Dyson index is $\beta=2$.

Although present understanding of the QCD phase diagram at intermediate $\mu_B$ 
is largely model-dependent \cite{Fukushima:2010bq}, 
the ground state of QCD at low $T$ and at sufficiently large $\mu_B$ 
is expected to be a color superconductor \cite{Alford:2007xm} based on the
controlled weak-coupling calculations.  
At sufficiently large $\mu_B$, the physics near the Fermi surface 
is described by the weakly interacting quarks due to the asymptotic freedom.
The perturbative one-gluon exchange interaction,
\beq
{\cal L}_{\rm OGE}=-G(\bar \psi \gamma^{\mu} T^{\SU}_a \psi)^2,
\eeq
is dominant at large $\mu_B$ with $G>0$.
The color factor of this interaction in the $\psi \psi$-channel is 
\beq
(T^{\SU}_a)_{\alpha \beta} (T^{\SU}_a)_{\gamma \delta}
= \frac{N_c-1}{N_c} (T^{\U}_S)_{\alpha \gamma} (T^{\U}_S)_{\delta \beta}
- \frac{N_c+1}{N_c} (T^{\SU}_A)_{\alpha \gamma} (T^{\SU}_A)_{\delta \beta},
\eeq
where $T^{\U}_S$ and $T^{\SU}_A$ are symmetric and antisymmetric 
$N_c \times N_c$ Hermitian matrices.
This is attractive in the color antisymmetric channel.
According to the Bardeen-Cooper-Schrieffer (BCS) mechanism, 
any infinitesimally small attractive interaction between quarks
leads to the condensation of quark-quark pairs, the diquark condensate.
Because the positive parity state is favored by instanton effects \cite{Alford:1997zt, Rapp:1997zu}, 
the pairing in the spin-parity $0^+$ channel is the most favorable energetically.
Considering the Pauli principle, the condensate must be flavor antisymmetric.
Therefore, the condensate $\langle \psi^T C\gamma_5 T_A \tau_{A'} \psi \rangle$
is formed at large $\mu_B$, where both $A$ and $A'$ denote antisymmetric representations. 
Since the diquark condensate breaks $\SU(N_c)$ color symmetry,
this is called the color superconductivity \cite{Alford:2007xm}.

\subsection{QCD: $\SU(N_c \geq 3)$ gauge theory at finite $\mu_I$ ($\beta=2$)}
Now we turn to QCD at finite $\mu_I$ and $\mu_B=0$ 
(see \cite{Son:2000xc} for details).
Due to the property $\tau_1 \tau_3 \tau_1 = -\tau_3$, 
the Dirac operator (\ref{eq:D_mu_I}) satisfies
\beq
\label{eq:anti-unitary2}
\tau_1 \gamma_5 {\cal D} \gamma_5 \tau_1 = {\cal D}^{\dag}.
\eeq
From this relation and the chiral symmetry $\{\gamma_5,{\cal D}\}=0$,
if we take an eigenvalue $i \lambda_n$ of the Dirac operator (\ref{eq:D_mu_I}),
eigenvalues appear in quartet 
$(i\lambda_n, -i\lambda_n, i\lambda_n^*, -i\lambda_n^*)$.
Therefore, one finds that $\det[{\cal D}(\mu)+m] \geq 0$
and the standard Monte-Carlo simulation technique can be used at finite $\mu_I$.

When $m = \mu_I = 0$, the Lagrangian has $\SU(2)_L \times \SU(2)_R$ symmetry. 
If the degenerate quark mass $m$ is turned on, 
the symmetry is explicitly broken to $\SU(2)_{L+R}$.
The isospin chemical potential $\mu_I$ further breaks $\SU(2)_{L+R}$
down to $\U(1)_{L+R}$. Note that this symmetry is different from $\U(1)_B$.

\begin{figure}[t]
\begin{center}
\includegraphics[width=10cm]{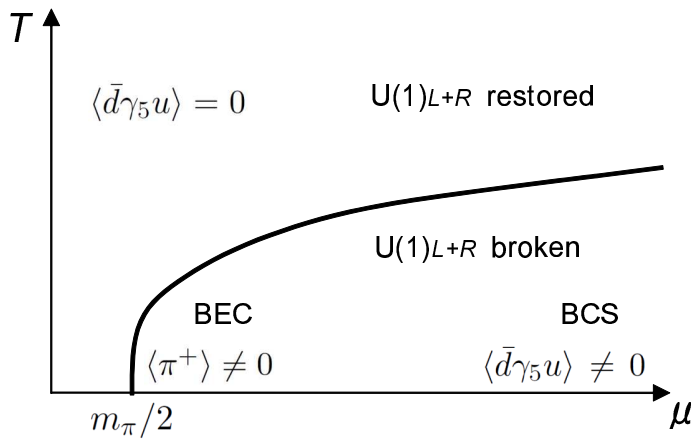}
\end{center}
\vspace{-0.5cm}
\caption{Phase diagram of QCD at finite $\mu_I=2\mu$.}
\label{fig:muI}
\end{figure}

Let us consider the zero-temperature ground state
at small $\mu_I$ and sufficiently large $\mu_I$
where the theory is analytically controllable.
Unlike the phase structure of QCD at finite $\mu_B$,
that of QCD at finite $\mu_I$ (with degenerate quark mass)
is rather well understood because of the absence of the sign problem 
and the constraints from the rigorous QCD inequalities 
\cite{Weingarten:1983uj, Witten:1983ut, Nussinov:1983hb, Espriu:1984mq}
\cite{Kogut:1999iv, Kogut:2000ek}, where
both properties follow from the relation (\ref{eq:anti-unitary2}).
The phase diagram in the $T$-$\mu$ plane is summarized in Fig.~\ref{fig:muI}.
For small $\mu_I$ below the $\rho$ meson mass $m_{\rho}$, 
we can concentrate on the pions at low-energy.
When $\mu_I < m_{\pi}$, no particles can be excited so that the ground state
is the same as the QCD vacuum at $\mu_I=0$.
On the other hand, when $\mu_I > m_\pi$, it is favorable to excite $\pi^+$ whose
excitation energy is $m_{\pi} - \mu_I < 0$; the ground state turns into 
the Bose-Einstein condensation (BEC) phase, $\langle \pi^+ \rangle \neq 0$,
where $\U(1)_{L+R}$ is spontaneously broken down to ${\mathbb Z}_2$.

At sufficiently large $\mu_I$, on the other hand, 
$\U(1)_{L+R}$ symmetry is spontaneously 
broken by the BCS-type diquark pairing.
The color factor of the one-gluon exchange interaction 
in the $\bar \psi \psi$-channel is
\beq
(T^{\SU}_a)_{\alpha \beta} (T^{\SU}_a)_{\gamma \delta}
=\frac{N_c^2-1}{2N_c^2}({\bf 1})^{N_c}_{\alpha \delta}({\bf 1})^{N_c}_{\gamma \beta}
-\frac{1}{N_c}(T^{\SU}_a)_{\alpha \delta} (T^{\SU}_a)_{\gamma \beta},
\eeq
which is attractive in the color-singlet channel.
The spin-parity $0^-$ channel is favored by the instanton effects, and
the condensate must be flavor antisymmetric from the Pauli principle,
$\langle \bar d \gamma_5 u \rangle \neq 0$. 
This is consistent with the requirement from the QCD inequality \cite{Son:2000xc}.
Note that, although the mechanism leading to the pairing is similar
to the color superconductivity in QCD at large $\mu_B$,
this condensate has different quantum numbers and does not break 
$\SU(N_c)$ color symmetry.

Since the BCS pairing $\langle \bar d \gamma_5 u \rangle \neq 0$ at large $\mu_I$ 
has the same quantum numbers and breaks the same $\U(1)_{L+R}$ symmetry 
as the BEC $\langle \pi^+ \rangle \neq 0$ at small $\mu_I$, it is natural to 
expect that the BEC and BCS regions are continuously connected 
without any phase transition (the BEC-BCS crossover).
At sufficiently high $T$, the condensate melts away and $\U(1)_{L+R}$ 
symmetry recovers. The critical temperature $T_c$ vanishes at $\mu=m_{\pi}/2$ 
and is expected to be an increasing function of $\mu$ in both BEC and BCS regions.

\subsection{$\SO(2N_c)$ gauge theory at finite $\mu_B$ ($\beta=4$)}
We next consider $\SO(2N_c)$ gauge theory at finite $\mu_B$. From the property
$(A^{\SO}_{\mu})^*=-A^{\SO}_{\mu}$, one has the relation: 
\beq
\label{eq:anti-unitary4}
C\gamma_5 {\cal D} C\gamma_5={\cal D}^*, \quad {\rm or} \quad [{\cal D}, C\gamma_5 K]=0,
\eeq
where $C$ is the charge conjugation matrix
satisfying $C^2=1$ and $C \gamma^{\mu} C = -\gamma^{\mu *}$ with all
$\gamma$ matrices being Hermitian, and $K$ is the complex conjugation operator.
The Dirac operator can be written in the form of quaternion real matrix [see (\ref{eq:Dyson4})]
and the Dyson index is $\beta=4$, the same universality class as 
$\SU(N_c)$ gauge theory with adjoint fermions \cite{Kogut:2000ek} (see also \cite{Zhang:2010kn}).
From (\ref{eq:anti-unitary4}) and the chiral symmetry 
$\{\gamma_5, {\cal D}\}=0$,
if $i\lambda_n$ is one of the eigenvalues of ${\cal D}$, eigenvalues appear in  
quartet $(i\lambda_n, -i\lambda_n, i\lambda_n^*, -i\lambda_n^*)$.
Note that, when $\lambda_n$ is real or pure imaginary, this quartet reduces to two sets of
doublets $(i\lambda_n, -i\lambda_n)$ with their eigenvectors being linearly independent
from the anti-unitary symmetry (\ref{eq:anti-unitary4}) \cite{Hands:2000ei}.
Therefore, $\det[{\cal D}(\mu)+m] \geq 0$ 
and the standard Monte-Carlo simulation technique is available at finite 
$\mu_B$ \cite{Cherman:2010jj, Cherman:2011mh}. 

When $m = \mu_B = 0$, the Lagrangian \eqref{eq:Lagrangian} has the
$\SU(N_{f})_{L}\times \SU(N_{f})_{R} \times \U(1)_{B} \times \U(1)_{A}$ 
symmetry at the classical level at first sight.  
However, chiral symmetry of the theory is enhanced to $\U(2N_{f})$
owing to the anti-unitary symmetry (\ref{eq:anti-unitary4}) 
\cite{Coleman:1980mx,Peskin:1980gc}. At the quantum level,
$\U(1)_A \subset \U(2N_{f})$ is explicitly broken by the axial anomaly
and $\SU(2N_f)$ symmetry remains.
One can actually rewrite the fermionic part of the Lagrangian (\ref{eq:Lagrangian})
manifestly invariant under $\SU(2N_f)$
using the new variable $\Psi=(\psi_L, \sigma_2 \psi_R^*)^T$:
\beq
{\cal L}_{\rm f}=i \Psi^{\dag} \sigma_{\mu} D_{\mu} \Psi,
\eeq
where $\sigma_{\mu}=(-i, \sigma_k)$ with the Pauli matrices $\sigma_k$.
The chiral symmetry $\SU(2N_f)$ is spontaneously broken down to 
$\SO(2N_{f})$ by the formation of the chiral condensate 
$\langle \bar{\psi}{\psi} \rangle$, leading to the $2N_f^2 + N_f -1$ Nambu-Goldstone (NG) bosons 
living on the coset space $\SU(2N_{f})/\SO(2N_{f})$.
In contrast to real QCD, there are not only $\U(1)_{B}$ neutral NG modes 
with the quantum numbers $\Pi_a=\bar{\psi} \gamma_{5} P_a \psi$ 
(just like the usual pions), but also $\U(1)_{B}$ charged NG modes with the quantum numbers 
$\Sigma_S = \psi^{T} C \gamma_5 Q_S \psi$ and 
$\Sigma_S^{\dag}= \psi^{\dag} C \gamma_5 Q_S \psi^*$.
Here $P_a$ are traceless and Hermitian $N_f \times N_f$ matrices, 
$P_a=P_a^{\dag}$ ($a=1,2,\cdots,N_f^2-1$), 
and $Q_S$ are symmetric $N_f \times N_f$ matrices, $Q_S^T=Q_S$
($S=1,2,\cdots,N_f(N_f + 1)/2$), in the flavor space.
The chiral perturbation theory describing these NG modes 
for small $\mu \lesssim m_{\rho}/2$ is exactly the same as that of $\SU(N_c)$ 
gauge theory with adjoint fermions at finite $\mu_B$ considered in \cite{Kogut:2000ek},
because their symmetry breaking patterns are the same:
the low-energy physics is dictated by the Dyson index $\beta=4$.

\begin{figure}[t]
\begin{center}
\includegraphics[width=10cm]{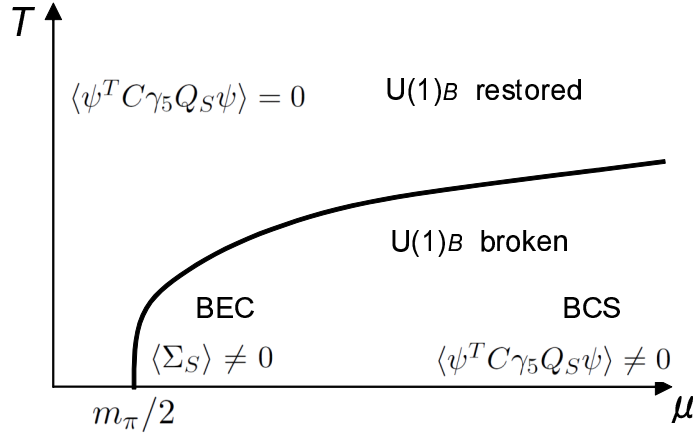}
\end{center}
\vspace{-0.5cm}
\caption{Phase diagram of $\SO(2N_c)$ gauge theory at finite $\mu_B$.}
\label{fig:SO}
\end{figure}

Let us consider the phase diagram of this theory.
For $m_{\pi}/2 < \mu \lesssim m_{\rho}/2$ at $T=0$, 
it is energetically favorable for the $\U(1)_B$ charged NG modes $\Sigma_S$ 
with the excitation energy $m_{\pi}-2\mu < 0$ to form the BEC, 
$\langle \Sigma_S \rangle \neq 0$.
On the other hand, at sufficiently large $\mu$, 
the one-gluon exchange interaction in the $\psi \psi$-channel
is attractive in the color symmetric channel,
leading to the condensation of the diquark pairing.
Due to the Pauli principle, the BCS diquark pairing must be flavor symmetric, 
and takes the form $\langle \psi^{T} C \gamma_5 Q_S \psi \rangle \neq 0$.
This BCS pairing has the same quantum numbers and breaks the same symmetry 
as the BEC at small $\mu_B$, 
and it is plausible that no phase transition occurs between 
the BEC and BCS regions. The phase diagram of this theory is similar 
to that of QCD at finite $\mu_I$, as shown in Fig.~\ref{fig:SO}.

\subsection{$\Sp(2N_c)$ gauge theory at finite $\mu_B$ ($\beta=1$)}
We turn to $\Sp(2N_{c})$ gauge theory at finite $\mu_B$. 
From the property $J_c A^{\Sp}_{\mu} J_c = (A^{\Sp}_{\mu})^*$ with
\beq
\label{eq:J_c}
J_c = -i\sigma_{2} \otimes \textbf{1}_{N_{c}}
\eeq 
one has the relation:
\beq
\label{eq:anti-unitary1}
J_c C\gamma_5 {\cal D} C\gamma_5 J_c = -{\cal D}^*, \quad
{\rm or} \quad [{\cal D}, i J_c C \gamma_5 K]=0.
\eeq
The Dirac operator can be written in the form of real matrix [see (\ref{eq:Dyson1})],
and the Dyson index is $\beta=1$, 
the same universality class as two-color QCD \cite{Kogut:1999iv, Kogut:2000ek}
(see also \cite{Zhang:2010kn}).
One then finds that $\det[{\cal D}(\mu)+m] \geq 0$ for even $N_f$.
Note here that even $N_f$ is necessary for the positivity 
unlike $\SO(2N_c)$ gauge theory, because the quartet structure 
of eigenvalues $(i\lambda_n, -i\lambda_n, i\lambda_n^*, -i\lambda_n^*)$
reduces to one set of doublet $(i\lambda_n, -i\lambda_n)$ when $\lambda_n$ is real 
or pure imaginary \cite{Hands:2000ei}.

When $m = \mu_B =0$, because of the anti-unitary symmetry (\ref{eq:anti-unitary1}),
chiral symmetry of the theory is enhanced to $\SU(2N_{f})$~\cite{Coleman:1980mx,Peskin:1980gc}.
This can be seen by rewriting the fermionic part of the Lagrangian (\ref{eq:Lagrangian})
using the new variable $\Psi=(\psi_L, \sigma_2 \tau_2 \psi_R^*)^T$:
\beq
{\cal L}_{\rm f}=i \Psi^{\dag} \sigma_{\mu} D_{\mu} \Psi,
\eeq
which is manifestly invariant under $\SU(2N_f)$.
The chiral symmetry is spontaneously broken down to 
$\Sp(2N_{f})$ by the formation of the chiral condensate 
$\langle \bar{\psi}{\psi} \rangle$, giving rise to $2N_f^2 - N_f -1$
NG bosons parametrized by the coset space $\SU(2N_{f})/\Sp(2N_{f})$. 
There are both $\U(1)_B$ neutral NG modes 
with the quantum numbers $\Pi_a=\bar{\psi} \gamma_{5} P_a \psi$ and 
$\U(1)_B$ charged NG modes with the quantum numbers 
$\tilde \Sigma_A = \psi^{T} C \gamma_5 J_c Q_A \psi$ and 
$\tilde \Sigma_A^{\dag}= \psi^{\dag} C \gamma_5 J_c Q_A \psi^*$.
Here $P_a$ are traceless and Hermitian $N_f \times N_f$ matrices, 
$P_a=P_a^{\dag}$ ($a=1,2,\cdots,N_f^2-1$), 
and $Q_A$ are antisymmetric $N_f \times N_f$ matrices, $Q_A^T=-Q_A$
($A=1,2,\cdots,N_f(N_f - 1)/2$), in the flavor space.
The chiral perturbation theory for small $\mu \lesssim m_{\rho}/2$ is 
the same as that of two-color QCD at finite $\mu_B$ \cite{Kogut:1999iv, Kogut:2000ek}
because of the same symmetry breaking pattern
dictated by the Dyson index $\beta=1$.

\begin{figure}[t]
\begin{center}
\includegraphics[width=10cm]{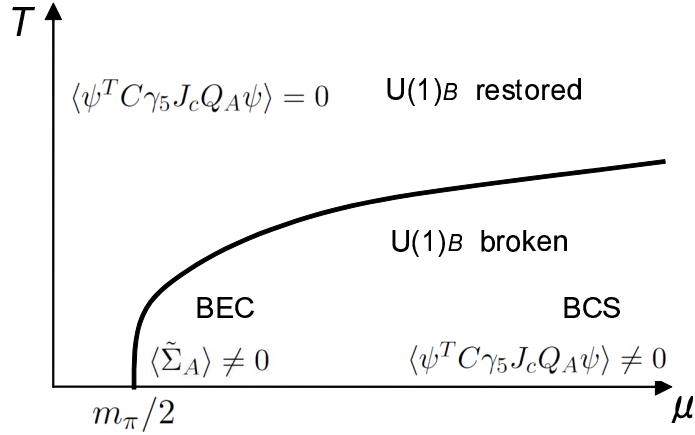}
\end{center}
\vspace{-0.5cm}
\caption{Phase diagram of $\Sp(2N_c)$ gauge theory at finite $\mu_B$.}
\label{fig:Sp}
\end{figure}

For $m_{\pi}/2 < \mu \lesssim m_{\rho}/2$ at $T=0$, the $\U(1)_B$ charged NG modes 
form the BEC, $\langle \tilde \Sigma_A \rangle \neq 0$. At sufficiently large $\mu$, 
the one-gluon exchange interaction in the $\psi \psi$-channel is 
attractive in the color antisymmetric channel
and induces the condensation of the color and flavor antisymmetric
BCS diquark pairing $\langle \psi^{T} C \gamma_5 J_c Q_A \psi \rangle \neq 0$.
Note that the diquark condensate has the different quantum numbers 
from the ones in $\SO(2N_c)$ gauge theory. The BEC-BCS crossover of 
the diquark pairing is expected in this theory again, 
as depicted in Fig.~\ref{fig:Sp}.

\subsection{Brief summary}
Before proceeding, we summarize the results in this section.

\begin{enumerate}
\item{QCD at finite $\mu_B$ (or $\mu_I$), $\SO(2N_c)$ gauge theory
at finite $\mu_B$, and $\Sp(2N_c)$ gauge theory at finite $\mu_B$ belong 
to the different universality classes denoted by the Dyson indices 
$\beta=2$, $\beta=4$, and $\beta=1$, respectively.}
\item{QCD at finite $\mu_I$ and $\SO(2N_c)$ and $\Sp(2N_c)$ gauge theories 
at finite $\mu_B$ have {\it no} fermion sign problem and
exhibit the BEC-BCS crossover phenomena in the phase diagrams.}
\end{enumerate}

Apparently, the phase diagrams of the three theories, QCD at finite $\mu_I$, 
$\SO(2N_c)$ gauge theory at finite $\mu_B$, 
and $\Sp(2N_c)$ gauge theory at finite $\mu_B$, resemble each other 
qualitatively.
In the next section, we will argue that these phase diagrams should be 
completely identical in the large-$N_c$ limit, 
including the chiral phase transition (not shown in figures).
Also we will show that the phase diagram of QCD at finite $\mu_B$
should also be identical to them outside the BEC-BCS crossover regions.

\section{Orbifold equivalence in the large-$N_c$ QCD and QCD-like theories}
\label{sec:QCD}
In this section, we first briefly recapitulate the basic idea 
of the orbifold equivalence in large-$N_c$ gauge theories.
We then construct orbifold projection from $\SO(2N_c)$ or 
$\Sp(2N_c)$ gauge theory at finite $\mu_B$ to $\SU(N_c)$ QCD 
at finite $\mu_B$ or $\mu_I$.
The relations between these theories via orbifold projections are summarized 
in Fig.~\ref{fig:QCD}.\footnote{We can also construct the orbifold projections
(not shown in Fig.~\ref{fig:QCD})
from $\SO(2N_c)$ gauge theory with adjoint fermions at finite $\mu_B$ to 
$\SU(N_c)$ QCD with adjoint fermions at finite $\mu_B$ and to
$\SU(N_c)$ QCD with antisymmetric fermions at finite $\mu_I$,
generalizing the argument of \cite{Kovtun:2003hr} to the case with 
finite chemical potential.
The resultant equivalence between $\SU(N_c)$ QCD with adjoint fermions
and $\SU(N_c)$ QCD with antisymmetric fermions is a generalization of 
the {\it orientifold equivalence} \cite{Armoni:2003gp} to finite chemical potential.}

\subsection{Perturbative orbifold equivalence in pure gauge theories}
\label{sec:pert}
The main idea of the orbifold projection is as follows: 
we identify a discrete subgroup of the symmetry group of the ``parent" theory, 
that is the $\SO(2N_{c})$ or $\Sp(2N_c)$ gauge theory in our case.
We then eliminate all of the degrees of freedom 
in the parent theory which are not invariant under the discrete symmetry.  
This gives the ``daughter" theory, which will turn out to be the $\SU(N_c)$ 
gauge theory.  
We use a $\mathbb{Z}_{2}$ subgroup of the original 
$\SO(2N_{c})$ or $\Sp(2N_c)$ gauge theory for the orbifold projection.
In the large-$N_c$ limit for fixed $N_f$ (the 't Hooft limit), 
correlation functions of operators ${\cal O}_i^{(p)}$ 
in the parent theory invariant under the projection symmetry
(which we call the ``neutral operators"), 
and those of the operators ${\cal O}_i^{(d)}$ in the daughter theory 
made up of the projected fields, coincide to all orders 
in the perturbation theory \cite{Bershadsky:1998cb}, 
\begin{eqnarray}
\label{eq:equivalence}
\langle{\cal O}_1^{(p)}{\cal O}_2^{(p)}\cdots\rangle_{p}
=
\langle{\cal O}_1^{(d)}{\cal O}_2^{(d)}\cdots\rangle_{d}.  
\end{eqnarray}
In this sense, the parent theory and the daughter theory are equivalent.
Especially, magnitudes of neutral order parameters characterizing the phases
should be exactly the same.

Let us first consider the orbifold projection of the pure $\SO(2N_c)$ gauge theory.
(For an earlier work of the orbifold projection from $\SO(2N_{c})$ to 
$\SU(N_{c})$, see \cite{Unsal:2006pj}.) 
For the ``projection," we use ${\mathbb Z}_4$ subgroup of $\SO(2N_c)$ generated by 
$J_c$ defined in (\ref{eq:J_c}). $J_c$ satisfies the condition, 
\begin{eqnarray}
\label{eq:regularity}
\tr (J_c^n) =0,
\end{eqnarray}
when $J_c^n$ does not belong to the center of $\SO(2N_c)$, i.e., $J_c^n\neq\pm\textbf{1}_{2N_c}$.
This condition is called the ``regularity condition," which is 
utilized in the perturbative proof of the orbifold equivalence.
The transformation of $A^{\SO}_\mu$ induced by $J_c$ is written as
\begin{eqnarray}
(A^{\SO}_\mu)_{ij}\to (J_c)_{ii'} (A^{\SO}_\mu)_{i'j'} (J_c)^{-1}_{j'j}, 
\end{eqnarray}
which constitutes ${\mathbb Z}_2$ subgroup of the gauge group. 
We define the projection condition for the gauge field $A^{\SO}_{\mu}$
to be invariant under the ${\mathbb Z}_2$ subgroup:
\begin{eqnarray}
\label{eq:projection_gauge}
(A^{\SO}_\mu)_{ij}= (J_c)_{ii'} (A^{\SO}_\mu)_{i'j'} (J_c)^{-1}_{j'j}. 
\end{eqnarray}
The gauge field $A^{\SO}_{\mu}$ satisfying this condition can be obtained 
by using the projector ${\cal P}$ defined as
\begin{eqnarray}
\label{eq:projection_factor}
{\cal P}(A^{\SO}_\mu)=\frac{1}{4}\sum_{n=0}^3 J_c^n 
A^{\SO}_\mu J_c^{-n}=\frac{1}{2}\left(A^{\SO}_\mu + J_c A^{\SO}_\mu J_c^{-1}\right),  
\end{eqnarray}
where $4$ is the number of elements of ${\mathbb Z}_4$. 

Remembering the property $(A^{\SO}_{\mu})^*=-A^{\SO}_{\mu}$, 
the gauge field $A^{\SO}_{\mu}$ can be written as
\begin{align}
A^{\SO}_\mu
=
i\left(
\begin{array}{cc}
A_\mu^A+B_\mu^A & C_\mu^A-D_\mu^S\\
C_\mu^A+D_\mu^S & A_\mu^A-B_\mu^A
\end{array}
\right),
\end{align}
where the fields $A_{\mu}^A$, $B_\mu^A$, and $C_\mu^A$ ($D_{\mu}^S$) 
are $N_{c} \times N_{c}$ anti-symmetric (symmetric) matrices.  
Under the $\mathbb{Z}_{2}$ symmetry, $A_\mu^A, D_\mu^S$ are even while $B_\mu^A, C_\mu^A$ are odd,  
so the orbifold projection sets $B_{\mu}^{A} = C_{\mu}^{A} = 0$. Hence we have
\begin{align}
A_{\mu}^{\rm proj}
=
i\left(
\begin{array}{cc}
A_{\mu}^A  & -D_\mu^S\\
D_\mu^S & A_\mu^A
\end{array}
\right).
\end{align} 
If one performs a unitary transformation in the color space using the matrix
\begin{eqnarray}
\label{eq:P_c}
P_c = \frac{1}{\sqrt{2}}\left(
\begin{array}{cc}
\textbf{1}_{N_{c}} & i \textbf{1}_{N_{c}} \\
\textbf{1}_{N_{c}} & -i \textbf{1}_{N_{c}}
\end{array} 
\right),
\label{unitary_transformation}
\end{eqnarray}
$A_{\mu}^{\rm proj}$ can be expressed by 
the $\U(N_{c})$ gauge field ${A}^{\U}_{\mu} \equiv D_{\mu}^{S} + i A^{A}_{\mu}$ as
\begin{eqnarray}
P_c A_{\mu}^{\rm proj} P_c^{-1} =
  \left(
\begin{array}{cc}
-({A}^{\U}_{\mu})^T & 0\\
0 & {A}^{\U}_{\mu}
\end{array} 
\right).
\label{gauge_field_diagonal_basis}
\end{eqnarray}
The top left component is the charge conjugation of the bottom right component,
$-({A}^{\U}_{\mu})^T=({A}^{\U}_{\mu})^C$.
At large $N_{c}$, we can neglect the difference between $\U(N_{c})$ and 
$\SU(N_{c})$ up to $1/N_{c}^{2}$ corrections, $A^{\U} \simeq A^{\SU}$.\footnote{Notice 
the chemical potential is introduced as a boundary condition for the $\U(1)$ part of 
the gauge field at infinity. However it is not easy to impose an appropriate 
boundary condition at infinity in lattice simulations; although the integration 
region of the path integral must be limited so that it does not alter the boundary 
condition (in particular constant shift of the gauge field, 
$A_{\mu} \rightarrow A_{\mu} + C$, is forbidden),
all field configurations, including constant shift, are summed over
in actual lattice simulations. Because of this lattice artifact, 
$\mu$-dependence disappears in $\U(N_c)$ gauge theory \cite{Barbour:1986jf}. 
In $\SU(N_c)$ gauge theory, this problem does not arise.}
Therefore, the action of the $\SO(2N_c)$ gauge theory,
\begin{eqnarray}
{\cal L}_{\SO} = \frac{1}{4 g_{\SO}^{2} } \tr (F^{\SO}_{\mu \nu})^2,
\end{eqnarray}
is projected to that of the $\SU(N_c)$ gauge theory,
\begin{eqnarray}
{\cal L}_{\rm proj} = \frac{2}{4 g_{\SU}^{2} } \tr ({F}^{\SU}_{\mu \nu})^2,
\end{eqnarray}
where ${F}^{\SU}_{\mu\nu}$ is the field strength of the $\SU(N_{c})$ gauge field 
${A}^{\SU}_{\mu}$. 
In this way, $\SU(N_c)$ gauge theory is obtained from $\SO(2N_c)$ gauge theory 
via the {\it orbifold projection}. 

We can similarly define the orbifold projection of the $\Sp(2N_c)$ gauge theory.
The symplectic algebra $\Sp(2N_c)$ formed by $2N_c\times 2N_c$ Hermitian matrices satisfies
\begin{eqnarray}
J_c A^{\Sp} + (A^{\Sp})^T J_c = 0,
\end{eqnarray}
and can be written as
\begin{align}
A^{\Sp}_\mu
=
\left(
\begin{array}{cc}
iA_\mu^A + B_\mu^S & C_\mu^S - iD_\mu^S\\
C_\mu^S + iD_\mu^S & iA_\mu^A - B_\mu^S
\end{array}
\right).
\end{align}
Here the fields $A_{\mu}^A$ ($B_\mu^S$, $C_\mu^S$, and $D_{\mu}^S$) 
are $N_c\times N_c$ anti-symmetric (symmetric) matrices.  
If we choose the same projection condition for $A^{\Sp}$ as (\ref{eq:projection_gauge}),
\beq
A^{\Sp}_{\mu} = J_c A^{\Sp}_{\mu} J_c^{-1}.
\eeq
one obtains $B_{\mu}^{S} = C_{\mu}^{S} = 0$. 
This gives the $\SU(N_c)$ gauge theory again.

The mapping rule of the large-$N_c$ orbifold equivalence is as follows:
if we equate the parent action with {\it twice} the daughter action 
for the ${\mathbb Z}_2$ orbifold projection \cite{Kovtun:2003hr,Kovtun:2004bz},
\beq
\label{eq:recipe}
{\cal L}_{\SO(\Sp)} \rightarrow 2 {\cal L}_{\SU},
\eeq
these two theories are equivalent in the sense that 
(\ref{eq:equivalence}) holds within the neutral sectors 
i.e., the correlation functions of the neutral operators.
The reason of this factor 2 will be clarified below [see (\ref{eq:factor})].
From the recipe (\ref{eq:recipe}), we must take the action of the daughter 
$\SU(N_c)$ gauge theory as
\begin{eqnarray}
{\cal L}_{\SU} = \frac{1}{4 g_{\SU}^{2} } \tr ({F}^{\SU}_{\mu \nu})^2,
\end{eqnarray}
with the coupling constant $g_{\SU}$ satisfying
\beq
g_{\SO(\Sp)}=g_{\SU}.
\eeq

Let us give a schematic of the proof of the orbifold equivalence
to all orders in the perturbation theory following Bershadsky and Johansen \cite{Bershadsky:1998cb}.
This can be easily generalized to the cases with the fundamental fermions \cite{Cherman:2010jj},
as we will see in the next subsection. 
We assume that the gauge fixing condition is consistent with the projection -- the ghosts are 
related by the projection, and all propagators in two theory takes the same form up to the color factors.

\begin{figure}[t]
\begin{center}
\includegraphics[width=6.5cm]{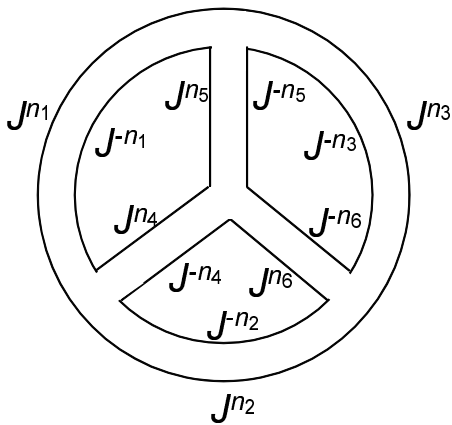}
\end{center}
\vspace{-0.5cm}
\caption{A vacuum planar diagram in the double-line notation.}
\label{fig:vacuum}
\end{figure}

As a pedagogical demonstration, consider a vacuum planar diagram
of the $\SO(2N_c)$ or $\Sp(2N_c)$ gauge theory in Fig.~\ref{fig:vacuum}.  
In order to obtain the $\SU(N_c)$ diagram from the $\SO(2N_c)$ or $\Sp(2N_c)$ diagram, 
we insert the projector ${\cal P}$ to each propagator.
Because the 't Hooft couplings are taken to be the same and the propagators are 
the same up to color factors, the only difference, if exists, comes from the contractions 
of color indices. Remembering (\ref{eq:projection_factor}), the additional 
kinematic factor multiplied by the $\SU(N_c)$ diagram is
\begin{eqnarray}
\label{eq:example}
\sum_{n_i=0,1}
\left(\frac{1}{2}\right)^{N_P}
\cdot \tr( J^{-n_1}J^{n_4}J^{n_5})
\cdot \tr( J^{-n_2}J^{-n_4}J^{n_6})
\cdot \tr( J^{-n_3}J^{-n_5}J^{-n_6})
\cdot \tr( J^{n_1}J^{n_2}J^{n_3}),
\nonumber\\
\end{eqnarray}
where $J=-i\sigma_2$ is a $2\times 2$ matrix and
the factor $(1/2)^{N_P}$ comes from the projectors
with $N_P=6$ being the number of propagators.  
Because of the regularity condition (\ref{eq:regularity}), 
it is nonvanishing only when 
\begin{eqnarray}
\label{eq:constraints}
& &
J^{-n_1}J^{n_4}J^{n_5}=\pm\textbf{1}_2,
\quad 
J^{-n_2}J^{-n_4}J^{n_6}=\pm\textbf{1}_2, 
\quad
J^{-n_3}J^{-n_5}J^{-n_6}=\pm\textbf{1}_2, 
\quad
J^{n_1}J^{n_2}J^{n_3}=\pm\textbf{1}_2. 
\nonumber\\
\end{eqnarray}
The last condition follows from the others, and hence, there are $N_L-1=3$ independent constraints, 
where $N_L=4$ is the number of color index loops. 
In (\ref{eq:example}) the combinations of $n_i=0,1$ for all $N_P=6$ propagators 
under the $N_L-1=3$ constraints give a factor of $2^{6-3}$, 
traces over color indices give $2^4$, and thus,
the total factor is 
\beq
2^{-6} \cdot 2^{6-3}\cdot 2^{4}=2.
\eeq 

Generally, for any given planar vacuum diagram with $N_P$ propagators and $N_L$ loops,
the projectors give a factor of $(1/2)^{N_P}$,
the combinations of $n_i=0,1$ ($i=1,2,\cdots,N_P$) 
under the $N_L-1$ constraints give $2^{N_P-(N_L-1)}$, and the traces over color indices 
give $2^{N_L}$. Therefore, the total factor is always the same:
\beq
\label{eq:factor}
2^{-N_P} \cdot 2^{N_P-(N_L-1)} \cdot 2^{N_L}=2.
\eeq
This factor 2 reflects the fact that the number of degrees 
of freedom in the parent theory is twice larger than that in the daughter theory.
Hence the vacuum energy per degree of freedom is identical between these theories.
This is why the recipe (\ref{eq:recipe}) is necessary to match the degrees of 
freedom between the parent and daughter theories.
However, this argument does not hold for nonplanar diagrams. Actually, one can
check that the number of independent constraints is no longer $N_L-1$, 
and the kinematic factor is different from 2 counted in (\ref{eq:factor}) \cite{Bershadsky:1998cb}.
This is why we need to take the large-$N_c$ limit to suppress the nonplanar diagrams.
We can repeat the same argument to any correlation functions 
of neutral operators (gauge-invariant operators in this case)
in any dimension and any gauge group. 

Note that this equivalence holds as long as the projection symmetry, $\mathbb{Z}_{2}$ 
subgroup of $\SO(2N_{c})\times \U(1)_{B}$ or $\Sp(2N_{c})\times \U(1)_{B}$, 
is not broken spontaneously \cite{Kovtun:2004bz}.
We will come to this issue in more detail in Sec.~\ref{sec:condition}.

\subsection{Perturbative orbifold equivalence with fermions}
\label{sec:projection}
In this subsection, we further include the fermions and 
define the orbifold projections from $\SO(2N_c)$ and $\Sp(2N_c)$ 
gauge theories to $\SU(N_c)$ gauge theory at finite $\mu_B$ or $\mu_I$.\footnote{The 
orbifold projection can be generalized to the case with both finite $\mu_B$ and finite $\mu_I$.
More generally, one can construct the orbifold projection in the system 
where each flavor has different chemical potential $\mu_f$ ($f=1,2,\cdots,N_f$).}

In order to obtain fermions at finite $\mu_B$, we use $\mathbb{Z}_{4}$ subgroup 
of $\SO(2N_{c})$ or $\Sp(2N_c)$ gauge group generated by $J_c$ and 
$\mathbb{Z}_{4}$ subgroup of $\U(1)_{B}$ generated by $\omega = e^{i \pi/2}$.
We choose the projection condition as
\beq
\label{eq:projection_baryon}
\psi^{\SO(\Sp)}_{a} = \omega (J_c)_{aa'} \psi^{\SO(\Sp)}_{a'},
\eeq
which generates $\mathbb{Z}_{2}$ subgroup of $\SO(2N_{c}) \times \U(1)_{B}$
or $\Sp(2N_{c}) \times \U(1)_{B}$.
The color $2N_c$-component fundamental fermion is decomposed into 
two $N_c$-component fields, 
\begin{eqnarray}
\psi^{\SO(\Sp)}
=
\left(
\begin{array}{c}
\psi_a\\
\psi_b
\end{array}
\right), 
\end{eqnarray}
with $a$ and $b$ being the color indices. 
Performing the unitary transformation $P_c$ defined in (\ref{eq:P_c}), 
we have 
\begin{eqnarray}
P_c\psi^{\SO(\Sp)}
=
\left(
\begin{array}{c}
\psi_+\\
\psi_-
\end{array}
\right), 
\end{eqnarray}
where $\psi_\pm \equiv (\psi_a \pm i\psi_b)/\sqrt{2}$.
The fermions $\psi_+$ and $\psi_-$, 
which couple to $(A_{\mu}^{\SU})^C$ and $A_{\mu}^{\SU}$
from \eqref{gauge_field_diagonal_basis},
transform as antifundamental and fundamental representations 
under $\SU(N_c)$, respectively. 
After the projection (\ref{eq:projection_baryon}), only the fermion $\psi_-$ survives. 
Taking into account the relation (\ref{eq:recipe}), the action of the daughter theory
is given by
\begin{eqnarray}
{\cal L}_{\SU} = \frac{1}{4 g_{\SU}^{2} } \tr ({F}^{\SU}_{\mu \nu})^2
+ 
\sum_{f=1}^{N_f}
\bar{\psi}^{\SU}_{f}\left( \gamma^{\mu} {D}_{\mu} + m + \mu \gamma^4\right)\psi^{\SU}_{f}, 
\end{eqnarray}
where $\psi^{\SU}_{a} = \sqrt{2} \psi_{-}$,  ${D}_{\mu} = \partial_{\mu} + i {A}^{\SU}_{\mu}$.
This theory is QCD at finite baryon chemical potential $\mu_B = N_c \mu$.

On the other hand, in order to obtain fermions at finite $\mu_I$ for even $N_f$, 
we use $J_c \in \SO(2N_c)$ [or $J_c \in \Sp(2N_c)$] and 
$J_i \in \SU(2)_{\rm isospin}$ defined by
\beq
\label{eq:J_i}
J_i = - i\sigma_2 \otimes 1_{N_f/2}.
\eeq
We choose the projection condition to be
\beq
\label{eq:projection_isospin}
(J_c)_{aa'} \psi_{a'f'}^{\SO(\Sp)} (J_i^{-1})_{f'f} = \psi_{af}^{\SO(\Sp)}. 
\eeq
The flavor $N_f$-component fundamental fermion is decomposed into 
two $(N_f/2)$-component fields, 
\begin{eqnarray}
\psi^{\SO(\Sp)}
=
(\psi_f \ \psi_g), 
\end{eqnarray}
with $f$ and $g$ being the flavor indices.
If we define $\varphi_{\pm}=(\psi_{\pm}^f \mp i \psi_{\pm}^g)/\sqrt{2}$
and $\xi_{\pm}=(\psi_{\pm}^f \pm i \psi_{\pm}^g)/\sqrt{2}$,
the fermions $\varphi_{\pm}$ survive but $\xi_{\pm}$ disappear after the projection
(\ref{eq:projection_isospin}).
Because $\varphi_{\pm}$ couple to $(A_{\mu}^{\SU})^C$ and $A_{\mu}^{\SU}$ respectively, 
the action of the daughter theory is expressed as
\begin{eqnarray}
{\cal L}_{\SU} = \frac{1}{4 g_{\SU}^{2} } \tr ({F}^{\SU}_{\mu \nu})^2
+ \sum_{\pm}
\bar{\psi}^{\SU}_{\pm}\left( \gamma^{\mu} {D}_{\mu} + m \pm \mu \gamma^4 \right)\psi^{\SU}_{\pm}, 
\end{eqnarray}
where $\psi^{\SU}_{+}=\sqrt{2}\varphi_-$ and $\psi^{\SU}_{-}=\sqrt{2}\varphi_+^C$.
This theory, in which ``$\pm$" can be
regarded as the isospin indices, 
has the isospin chemical potential $\mu_I=2\mu$.

Given the orbifold projections, we now prove the orbifold equivalence
of the gauge theories with fermions at finite chemical potential.
Let us consider a diagram (a correlation function or an observable)
in $\SO(2N_c)$ or $\Sp(2N_c)$ gauge theory which have quark loop(s).
Here we write both color and flavor index lines for each quark line
since the contractions of flavor index loops also give kinematical factors.

We first look into the case with finite $\mu_I$. 
Because the matrix $J_c$ multiplied from the right in (\ref{eq:projection_gauge}) 
is just replaced by $J_i$ in (\ref{eq:projection_isospin}),
we can repeat the proof for the pure gauge theory straightforwardly.
For a diagram with $N_P$ propagators and $N_L^{(c)}$ color ($N_L^{(f)}$ flavor) index loops,
we count additional kinematical factors originating from contractions of 
both colors and flavors that are multiplied by the parent $\SU(N_c)$ diagram.  
The projectors for the propagators give a factor of $(1/2)^{N_P}$, 
the combinations of $n_i=0,1$ ($i=1,2,\cdots,N_P$) under
the $N_L^{(c)}+N_L^{(f)} -1$ constraints give $2^{N_P-(N_L^{(c)} + N_L^{(f)} -1)}$, and
the traces over color and flavor indices give $2^{N_L^{(c)}}$ and $2^{N_L^{(f)}}$
respectively. Therefore, the total factor is
\beq
2^{-N_P} \cdot 2^{N_P - (N_L^{(c)} + N_L^{(f)} -1)} \cdot 2^{N_L^{(c)}} \cdot 2^{N_L^{(f)}}=2,
\eeq
for any number of quark loops $N_L^{(f)}$.

On the other hand, this argument is not applicable to the case with 
finite $\mu_B$ since the structure of the projection condition
(\ref{eq:projection_baryon}) is different from (\ref{eq:projection_gauge})
and (\ref{eq:projection_isospin}).
However we can still justify the orbifold equivalence as long 
as the number of quark loop is one.
Consider a diagram with $N_P$ propagators and $N_L^{(c)}$ color index loop and 
one fermion index loop. 
The fermion index loop gives no condition for $J^{n_i}$ in Eq.~(\ref{eq:constraints}),
and the number of the constraints for $n_i=0,1$ ($i=1,2,\cdots,N_P$)
is $N_L^{(c)}$. Then one again concludes that the total kinematical factor is 2. 
Note that this does not hold any more if the number of fermion index loop 
is more than one.

Therefore, remembering that the fermion loops are suppressed by a factor of $N_f/N_c$, 
the orbifold equivalence at finite $\mu_B$ holds only 
in the 't Hooft limit (large-$N_c$ limit for fixed $N_f$) 
while that at finite $\mu_I$ holds both in the 't Hooft and 
the Veneziano limits (large-$N_c$ limit for fixed $N_f/N_c$).

One can also argue to what order of $1/N_c$ the orbifold equivalence is valid. 
The leading corrections to the 't Hooft limit come from the
one-fermion-loop planar diagrams, which, as we have shown above,
do not distinguish between $\mu_B$ and $\mu_I$.
Therefore, the difference of the expectation values of gluonic operators,
such as the Polyakov loop, between QCD at finite $\mu_B$ and $\mu_I$ 
is at most $\sim (N_f/N_c)^2$ due to a two-fermion-loop planar diagram. 
In particular, the difference of the critical temperature of the 
deconfinement is $\sim (N_f/N_c)^2$, as previously discussed in 
\cite{Toublan:2005rq} without using the orbifold equivalence.

\subsection{Condition for the orbifold equivalence and the BEC-BCS crossover region}
\label{sec:condition}
From the orbifold projections we have constructed, the expectation values 
of the neutral operators for given $T$ and $\mu$ should be 
identical between the parent and daughter theories.
Note here that, in order for the orbifold equivalence to QCD at finite $\mu_B$ to hold, 
the projection symmetry, $\mathbb{Z}_{2}$ subgroup of the 
$\SO(2N_{c})\times \U(1)_{B}$ for the $\SO(2N_c)$ gauge theory, or
$\mathbb{Z}_{2}$ subgroup of the 
$\Sp(2N_{c})\times \U(1)_{B}$ for the $\Sp(2N_c)$ gauge theory,
must not be broken spontaneously \cite{Kovtun:2004bz}. 
However, the $\U(1)_B$ symmetry, whose ${\mathbb Z}_4$ subgroup is 
used for the projection to QCD at finite $\mu_B$, 
is broken down to ${\mathbb Z}_2$ inside the BEC-BCS crossover regions in 
figures~\ref{fig:SO} and \ref{fig:Sp}.
Therefore, the orbifold equivalence to QCD at finite $\mu_B$ is valid only 
outside the BEC-BCS crossover regions.
Note that, even at large $\mu$, above the critical temperature of the superfluid, 
$T>T_c$, the condensates melt away and the equivalence recovers.

On the other hand, the equivalence to QCD at finite $\mu_I$ is valid as long as 
the projection symmetry, $\mathbb{Z}_{2}$ subgroup of $\SO(2N_{c}) \times \SU(2)_F$
or $\mathbb{Z}_{2}$ subgroup of $\Sp(2N_{c}) \times \SU(2)_F$, is not broken. 
This condition is satisfied for arbitrary $\mu$ and $T$ in all the three theories
with flavor symmetry, QCD at finite $\mu_I$ and $\SO(2N_c)$ and 
$\Sp(2N_c)$ gauge theories at finite $\mu_B$. As a result, 
the chiral condensate $\langle \bar \psi \psi \rangle$, the superfluid gap $\Delta$, 
and their critical temperatures are identical; the phase diagrams characterized 
by these order parameters completely coincide in the large-$N_c$ limit.

In particular, QCD at finite $\mu_B$ is equivalent to QCD at finite $\mu_I$ 
outside the BEC-BCS crossover region.
Remembering that dropping the complex phase of the fermion determinant of 
QCD at finite $\mu_B$ reduces to QCD at finite $\mu_I$ for $N_f=2$,
this means that {\it the phase-quenched approximation in this region 
is exact at large-$N_c$ for neutral operators, e.g., the chiral condensate.}
This explains why the phase-quenched approximation works in QCD at high temperature, 
as already observed in the lattice QCD simulations for small $\mu$ \cite{Allton:2002zi}
and in model calculations such as the chiral random matrix model \cite{Klein:2003fy}, 
Nambu--Jona-Lasinio model \cite{Toublan:2003tt, Barducci:2005ut},
and hadron resonance gas model \cite{Toublan:2004ks}.
As we will show in sections~\ref{sec:RMT_equivalence} and \ref{sec:solveRMT}. 
the results of the chiral random matrix model can also be understood 
as a consequence of the orbifold equivalence in this model.

\subsection{Nonperturbative orbifold equivalence at high density}
\label{sec:nonpert}
In Sec.~\ref{sec:pert}, we discussed the orbifold equivalence at the perturbative level
in the large $N_c$ limit. One might ask if the equivalence does hold
nonperturbatively and to what extent the $1/N_c$ corrections are important.
For $\SU(N_c)$ QCD at large $\mu_I$ and $\SO(2N_c)$ and $\Sp(2N_c)$ gauge theories 
at large $\mu_B$, we can answer both questions for several quantities explicitly.
This is because the coupling constants $g_{\SU}$, $g_{\SO}$, and $g_{\Sp}$ are small 
due to the asymptotic freedom and the calculations are under theoretical control. 
Note that the weak-coupling does not necessarily mean that calculations are 
perturbative: as shown in (\ref{eq:gap}), (\ref{eq:SUgap}), (\ref{eq:SOgap}), 
and (\ref{eq:Spgap}) below, 
the coupling constant dependences of the BCS gaps are indeed nonperturbative.
Although our calculations do not constitute the proof of nonperturbative 
orbifold equivalence for all the neutral observables, 
they provide a nontrivial piece of evidence for the equivalence.

Let us consider $N_f=2$.
First of all, the chiral condensate $\langle \bar \psi \psi \rangle$ is 
vanishing in all the theories at asymptotically large $\mu$, and the equivalence is trivially satisfied.
The nontrivial quantity we can compare is the superfluid gap $\Delta$.
Its equivalence is required, e.g., from the equivalence of 
the fermion occupation number
$\langle \psi_{af}^{\dag}(t,{\bf x}) \psi_{bg}(t,{\bf y}) \rangle$.
The gap $\Delta$ can be computed using the technique in \cite{Son:1998uk}. 
The main modification in the gap equations compared with \cite{Son:1998uk} is the 
group theoretical factor involving $N_c$ given by
\beq
\label{eq:color_SU_B}
(T^{\SU}_a)_{\alpha \beta}^T (T^{\SU}_A)_{\beta \gamma} (T^{\SU}_a)_{\gamma \delta} 
&=&-\frac{N_c+1}{2N_c}(T^{\SU}_A)_{\alpha \delta}, \\
\label{eq:color_SU}
(T^{\SU}_a)_{\alpha \beta} (\textbf{1})^{N_c}_{\beta \gamma} (T^{\SU}_a)_{\gamma \delta} 
&=&\frac{N_c^2-1}{2N_c}(\textbf{1})_{\alpha \delta}^{N_c}, \\
\label{eq:color_SO}
(T^{\SO}_a)_{\alpha \beta}^T (\textbf{1})^{2N_c}_{\beta \gamma} (T^{\SO}_a)_{\gamma \delta} 
&=&-\frac{2N_c-1}{4}(\textbf{1})_{\alpha \delta}^{2N_c}, \\
\label{eq:color_Sp}
(T^{\Sp}_a)_{\alpha \beta}^T (J_c)^{2N_c}_{\beta \gamma} (T^{\Sp}_a)_{\gamma \delta} 
&=&-\frac{2N_c+1}{4}(J_c)_{\alpha \delta}^{2N_c},
\eeq
respectively. Solving the gap equations,
we obtain the following BCS gap (up to prefactor) for each theory:
\beq
\label{eq:gap}
\Delta^{\SU}_{\mu_B} & \sim & \mu \exp \left({-\frac{\pi^2}{g_{\SU}}\sqrt{\frac{6N_c}{N_c + 1}}}\right),
\\
\label{eq:SUgap}
\Delta^{\SU}_{\mu_I} & \sim & \mu \exp \left({-\frac{\pi^2}{g_{\SU}}\sqrt{\frac{6N_c}{N_c^2-1}}}\right),
\\
\label{eq:SOgap}
\Delta^{\SO}_{\mu_B} & \sim & \mu \exp \left({-\frac{\pi^2}{g_{\SO}}\sqrt{\frac{12}{2N_c-1}}}\right),
\\
\label{eq:Spgap}
\Delta^{\Sp}_{\mu_B} & \sim & \mu \exp \left({-\frac{\pi^2}{g_{\Sp}}\sqrt{\frac{12}{2N_c+1}}}\right).
\eeq
Equation (\ref{eq:gap}) is the result obtained in \cite{Schafer:1999fe}.
For $N_c=3$, $\Delta^{\SU}_{\mu_B}$ and $\Delta^{\SU}_{\mu_I}$ reduce 
to the results obtained in \cite{Son:1998uk} and \cite{Son:2000xc}, respectively.
Note that in the 't Hooft limit 
$\Delta^{\SU}_{\mu_I}$, $\Delta^{\SO}_{\mu_B}$, and $\Delta^{\Sp}_{\mu_B}$ 
remain finite (and the BEC-BCS crossover regions of these theories do not disappear)
while $\Delta^{\SU}_{\mu_B}$ is vanishing. 
This originates from the fact that the diagrams of the one-gluon exchange
responsible for $\Delta^{\SU}_{\mu_I}$, $\Delta^{\SO}_{\mu_B}$, and $\Delta^{\Sp}_{\mu_B}$ are planar,
whereas it is nonplanar for $\Delta^{\SU}_{\mu_B}$ \cite{Deryagin:1992rw, Shuster:1999tn}. 
These consequences are consistent with our claim in Sec.~\ref{sec:condition} that the orbifold equivalence holds 
between QCD at large $\mu_I$ and $\SO(2N_c)$ and $\Sp(2N_c)$ gauge theories at large $\mu_B$
while the equivalence with QCD at large $\mu_B$ does not hold inside the BEC-BCS crossover region.
This gives a simple example in QCD that, if the projection symmetry is broken,
the orbifold equivalence is not valid.

Now let us compare $\Delta^{\SU}_{\mu_I}$, $\Delta^{\SO}_{\mu_B}$, and $\Delta^{\Sp}_{\mu_B}$.
For the comparisons to the leading order, 
we take the ratios of the factors in the exponential between (\ref{eq:SUgap}), 
(\ref{eq:SOgap}), and (\ref{eq:Spgap}).
Remembering $g_{\SU}=g_{\SO}=g_{\Sp}$, the ratios read
\beq
\label{eq:SO/SU}
\alpha_{\SO/\SU}(N_c) 
= \sqrt{\frac{2(N_c^2-1)}{N_c(2N_c - 1)}}
= \left\{ {\begin{array}{*{20}c}
   1.033 \quad &(N_c=3) \\
   1 \quad &(N_c=\infty) \\
\end{array}} \right. , 
\\
\label{eq:Sp/SU}
\alpha_{\Sp/\SU}(N_c) 
= \sqrt{\frac{2(N_c^2-1)}{N_c(2N_c + 1)}}
= \left\{ {\begin{array}{*{20}c}
   0.873 \quad &(N_c=3) \\
   1 \quad &(N_c=\infty) \\
\end{array}} \right. .
\eeq
Clearly, the equivalence holds in the large-$N_c$ limit
between QCD at large $\mu_I$ and $\SO(2N_c)$ and $\Sp(2N_c)$ gauge theories
at large $\mu_B$.
It also turns out that the orbifold equivalence is rather well satisfied
even in real QCD, $N_c=3$.
For the complete equivalence of the BCS gap, one has to check if
the prefactors in (\ref{eq:SUgap}), (\ref{eq:SOgap}), and (\ref{eq:Spgap}) 
are also identical, for which subleading effects are important.
Here we simply assume the equivalence of the prefactor in the large-$N_c$ limit, 
and see whether the equivalence of other nonperturbative quantities follow or not.

Provided ${\Delta^{\SO}_{\mu_I}}/{\Delta^{\SU}_{\mu_B}} \rightarrow 1$
and ${\Delta^{\Sp}_{\mu_I}}/{\Delta^{\SU}_{\mu_B}} \rightarrow 1$
in the large-$N_c$ limit, the critical temperatures $T_c$ of the superfluid phases  
also coincide with each other, since $T_c$ is proportional to the BCS gap with
the universal proportionality factor:
\beq
T_c = \frac{e^{\gamma}}{\pi} \Delta,
\eeq
where $\gamma \approx 0.577$ is the Euler-Mascheroni constant.

Moreover we compute the magnitudes of the diquark and pion condensates 
following \cite{Schafer:1999fe} as
\beq
\langle \bar d \gamma_5 u \rangle^{\SU}_{\mu_I}
&=&d^{\SU}_{\mu_I}, \qquad \qquad \qquad \
|d^{\SU}_{\mu_I}|=2\sqrt{\frac{6N_c}{N_c^2-1}} \frac{\mu^2 \Delta^{\SU}_{\mu_I}}{\pi g_{\SU}},
\\
\langle \psi^f_a C \gamma_5 \psi^g_b \rangle^{\SO}_{\mu_B}
&=& \delta_{ab} \delta^{fg} d^{\SO}_{\mu_B}, \qquad \quad \ \
|d^{\SO}_{\mu_B}|=2\sqrt{\frac{12}{2N_c-1}} \frac{\mu^2 \Delta^{\SO}_{\mu_B}}{\pi g_{\SO}},
\\
\langle \psi^f_a C \gamma_5 \psi^g_b \rangle^{\Sp}_{\mu_B}
&=& (J_c)_{ab} (J_i)^{fg} d^{\Sp}_{\mu_B}, \quad \
|d^{\Sp}_{\mu_B}|=2\sqrt{\frac{12}{2N_c+1}} \frac{\mu^2 \Delta^{\Sp}_{\mu_B}}{\pi g_{\Sp}},
\eeq
where $J_c$ and $J_i$ are defined in (\ref{eq:J_c}) and (\ref{eq:J_i}).
The ratios read
\beq
\frac{|d^{\SO}_{\mu_B}|}
{|d^{\SU}_{\mu_I}|}
&=&\alpha_{\SO/\SU}(N_c) \frac{\Delta^{\SO}_{\mu_B}}{\Delta^{\SU}_{\mu_I}},
\\
\frac{|d^{\Sp}_{\mu_B}|}
{|d^{\SU}_{\mu_I}|}
&=&\alpha_{\Sp/\SU}(N_c) \frac{\Delta^{\Sp}_{\mu_B}}{\Delta^{\SU}_{\mu_I}},
\eeq
where $\alpha_{\SO/\SU}(N_c)$ and $\alpha_{\Sp/\SU}(N_c)$ are 
the quantities defined in (\ref{eq:SO/SU}) and (\ref{eq:Sp/SU}).
Both indeed approach unity in the large-$N_c$ limit when
${\Delta^{\SO}_{\mu_B}}/{\Delta^{\SU}_{\mu_I}} \rightarrow 1$
and ${\Delta^{\Sp}_{\mu_B}}/{\Delta^{\SU}_{\mu_I}} \rightarrow 1$.

Although we did not attempt in this paper, 
one should also be able to check the equivalence 
for other quantities at large $\mu$, such as 
the four-quark condensate $\langle (\bar \psi \psi) (\bar \psi \psi) \rangle$,
the pion decay constant $f_{\pi}$, and so on.

\subsection{Brief summary}
We summarize our results in this section.

\begin{enumerate}
\item{The whole phase diagrams described by the chiral condensate and the superfluid gap
should be universal in the large-$N_c$ limit between
QCD at finite $\mu_I$ and $\SO(2N_c)$ and $\Sp(2N_c)$ gauge theories at finite $\mu_B$.}
\item{The phase diagram of QCD at finite $\mu_B$ should also be identical to 
those of other theories above outside the BEC-BCS crossover regions. 
In particular, the phase-quenched approximation for the chiral condensate 
is exact in this region.}
\item{At asymptotically large chemical potentials, the equivalence is rather 
well satisfied even for $N_c=3$. 
From this fact, the phase-quenched approximation for the chiral condensate 
is expected to work well in real QCD.}
\end{enumerate}

\section{Orbifold equivalence in the chiral random matrix theories}
\label{sec:RMT}
If the orbifold equivalence holds between the original gauge theories,
it is natural to expect that the equivalence 
should hold at the level of the corresponding low-energy effective theories.
In this section, we show that the orbifold equivalence holds perturbatively
as well as nonperturbatively in the chiral random matrix theory (RMT), 
a solvable effective theory of QCD (or QCD-like theory) 
first introduced in \cite{Shuryak:1992pi}.
Actually the RMT has the size of the matrix $N$, 
which is taken to infinity (thermodynamic limit) in the end.
In this sense, the RMT is a ``large-$N$" matrix model, and hence, 
the perturbative proof of the orbifold equivalence given in Sec.~\ref{sec:pert} is 
applicable. 
Note that the size of the random matrix is not related to the number of color $N_c$.
The orbifold equivalence will be verified nonperturbatively in Sec~\ref{sec:solveRMT} 
by solving the RMT following \cite{Klein:2003fy, Klein:2004hv}.

\subsection{Chiral random matrix theories}
In this subsection, we briefly review the basic aspects of the RMT.
For reviews of the RMT in more detail,
see \cite{Verbaarschot:2000dy, Akemann:2007rf}.

\subsubsection{Chiral random matrix theories at small chemical potential}
\label{sec:lowRMT}
The partition function of the RMT is given by an integral over a Gaussian 
random matrix ensemble, instead of the average over the gauge field of the 
original Yang-Mills action:
\beq
\label{eq:RMT}
Z=\int d\Phi \prod_{i=1}^{N_f} \det {\cal D} \ e^{-\frac{N \beta}{2}G^2 \tr \Phi^{\dagger} \Phi},
\eeq
where $\Phi$ is an $N \times (N+\nu)$ random matrix element,
$N$ is the size of the system, and $\nu$ is the topological charge. 
We also introduced a suitable normalization with the parameter $G$ in the Gaussian.
Note that there is no spacetime coordinate in the theory; 
the size of the matrix $N$ corresponds to the spacetime volume.
It is taken to infinity in the end, corresponding to the thermodynamic limit.
 
The matrix structure of the Dirac operator ${\cal D}$ is chosen 
such that it has the correct anti-unitary symmetries and 
it reproduces the correct global symmetry breaking pattern of the system.
In particular, anti-Hermiticity and chiral symmetry of the Dirac operator 
at $\mu=m=0$ require
\beq
\label{eq:constraint}
{\cal D}^{\dagger}=-{\cal D}, \qquad \{{\cal D}, \gamma_5 \}=0.
\eeq
The quark mass $m$, quark chemical potential $\mu$ 
\cite{Stephanov:1996ki}, and temperature $T$ \cite{Halasz:1998qr} can be 
incorporated into ${\cal D}$.
While $T$ does not destroy any relation in (\ref{eq:constraint}),
$\mu$ and $m$ break the former and the latter relations in (\ref{eq:constraint}),
respectively.

Depending on the anti-unitary symmetries of the Dirac operator,
the ensemble is distinguished with the real, complex, or quaternion real 
[see (\ref{eq:quaternion}) for the definition] matrices
denoted by the Dyson index $\beta=1$, $\beta=2$, and $\beta=4$, respectively.
The value of $\beta$ corresponds to the degrees of freedom per matrix element.
QCD or QCD-like theory in each universality class and the corresponding RMT
is listed as follows \cite{Halasz:1997fc}:
\begin{itemize}
\item{
The Dirac operator of $\SU(N_c \geq 3)$ QCD has no anti-unitary symmetry
and the corresponding RMT belongs to $\beta=2$.
The Dirac operator is taken as 
\begin{eqnarray}
\label{eq:Dyson2}
{\cal D}
=
\left(
\begin{array}{cc}
m_f \textbf{1}& \Phi+\mu\textbf{1} \\
-\Phi^\dagger +\mu\textbf{1} & m_f \textbf{1}
\end{array}
\right),
\end{eqnarray}
where $\Phi$ is an $ N\times (N + \nu)$ complex matrix and $m_f$ ($f=1,2,\cdots,N_f$)
are the quark masses.}

\item{
The Dirac operators of $\SU(N_c=2)$ QCD and $\Sp(2N_c)$ gauge theory
have the anti-unitary symmetries,
${\cal D} T_2^{\SU} \gamma_5 C = T_2^{\SU} \gamma_5 C {\cal D}^*$ and 
${\cal D} iJ_c \gamma_5 C = iJ_c \gamma_5 C {\cal D}^*$, respectively. 
Here $C$ is the charge conjugation matrix, $T_2^{\SU}$ is the
antisymmetric generator of $\SU(2)$, and $J_c$ is defined in (\ref{eq:J_c}).
The corresponding RMT belongs to the universality class $\beta=1$ and
the Dirac operator is taken as 
\begin{eqnarray}
\label{eq:Dyson1}
{\cal D}
=
\left(
\begin{array}{cc}
m_f \textbf{1}& \Phi+\mu\textbf{1} \\
-\Phi^T +\mu\textbf{1} & m_f \textbf{1}
\end{array}
\right),
\end{eqnarray}
where $\Phi$ is an $ N\times (N + \nu)$ real matrix.}

\item{
The Dirac operators of $\SU(N_c)$ QCD with adjoint fermions and 
$\SO(N_c)$ gauge theory have the anti-unitary symmetry, 
${\cal D} \gamma_5 C = \gamma_5 C{\cal D}^*$.
The corresponding RMT belongs to the universality class $\beta=4$ 
and the Dirac operator is taken as
\begin{eqnarray}
\label{eq:Dyson4}
{\cal D}
=
\left(
\begin{array}{cc}
m_f \textbf{1} & \Phi+\mu\textbf{1} \\
-\Phi^\dagger +\mu\textbf{1} & m_f \textbf{1}
\end{array}
\right),   
\end{eqnarray}
where $\Phi$ is a $2N \times 2(N + \nu)$ quaternion real matrix 
[see (\ref{eq:quaternion}) for the definition].}
\end{itemize}
These RMTs at finite $\mu$ can alternatively be formulated by 
the two-matrix representation \cite{Osborn:2004rf, Akemann:2005fd} 
where the identity matrix multiplied by $\mu$ is replaced by the random matrix element.
The effect of temperature $T$ can be incorporated as the (first) 
Matsubara frequencies by changing $\mu \rightarrow \mu + iT$ 
for one half of the determinant and $\mu \rightarrow \mu - iT$ 
for the other half of the determinant in the simplest model 
\cite{Halasz:1998qr}.\footnote{Introducing $T$ in this way
may break flavor symmetry of the system in some cases. However, the final result 
is shown to be equivalent to that of the correct prescription preserving 
flavor symmetry \cite{Vanderheyden:2005ux}.}

There is a regime (called the $\epsilon$-regime) where the RMT is exactly equivalent to QCD: 
when the typical scale of the system $L$ is much smaller than the pion Compton wavelength 
and is much larger than the inverse of the scale of chiral symmetry breaking \cite{Leutwyler:1992yt},
\beq
\label{eq:low_epsilon}
\frac{1}{\Lambda_{\chi}} \ll L \ll \frac{1}{m_{\pi}}, \qquad \mu L \ll 1,
\eeq
QCD reduces to a theory of zero momentum modes of pions.
In this regime, the system has a universality, i.e., the dynamics depends 
only on the symmetry breaking pattern and is independent of the microscopic details;
QCD can be replaced by the RMT with the same global symmetry breaking pattern.
Outside the $\epsilon$-regime, the universality is lost. However, the RMT is 
still useful as a schematic model to study the qualitative properties of QCD 
such as the phase structure at finite $T$ and $\mu$ \cite{Halasz:1998qr}.
The advantage of the RMT is that it can be solved analytically 
although QCD itself cannot be.

\subsubsection{Chiral random matrix theories at large chemical potential}
\label{sec:highRMT}
Recently, a new class of RMTs which describe the superfluid phase at asymptotically large $\mu$ 
have been identified in \cite{Kanazawa:2009en}.\footnote{The mathematical aspects of the 
same two-matrix model for $\beta=1$ were previously studied in 
\cite{Akemann:2009, Akemann:2009fc}.} 
The partition function is given by
\beq
Z=\int dA dB \prod_{i=1}^{N_f} \det {\cal D} \ e^{- \frac{N\beta}{2}G^2 \tr (A^{\dagger} A + B^{\dagger} B)},
\eeq
where $A$ and $B$ are $N \times N$ spacetime independent random matrix elements.
Here only the topological sector $\nu=0$ is considered because the topological susceptibility 
is strongly suppressed at large $\mu$ \cite{Schafer:2002ty, Yamamoto:2008zw}. 
Chiral symmetry 
\beq
\{{\cal D}, \gamma_5 \}=0,
\eeq 
is preserved at finite $\mu$ in the chiral limit $m=0$, 
but anti-Hermiticity of the Dirac operator ${\cal D}^{\dagger}=-{\cal D}$ is lost.
The non-Hermitian Dirac operator is taken as 
\begin{eqnarray}
{\cal D}
=
\left(
\begin{array}{cc}
m_f \textbf{1}_{N}& A_{N} \\
B_{N} & m_f \textbf{1}_{N}
\end{array}
\right),
\end{eqnarray}
where both $A$ and $B$ are the real, complex, or quaternion real matrices,
denoted by the Dyson index $\beta=1$, $\beta=2$, and $\beta=4$, respectively.
The $\beta=1$ RMT corresponds to $\SU(2)$ QCD \cite{Kanazawa:2009en} 
and $\Sp(N_c)$ gauge theory at large $\mu_B$, 
$\beta=2$ RMT to $\SU(N_c \geq 3)$ QCD at large $\mu_I$, 
and $\beta=4$ RMT to adjoint QCD and $\SO(2N_c)$ gauge theory at large $\mu_B$.

We can define the $\epsilon$-regime at large $\mu$ where the system has
the universality and QCD is equivalent to the RMT \cite{Yamamoto:2009ey, Kanazawa:2009ks}:
\beq
\label{eq:high_epsilon}
\frac{1}{\Delta} \ll L \ll \frac{1}{m_{\pi}},
\eeq
where $\Delta$ is the BCS gap and $m_{\pi}$ is the pion mass associated with chiral symmetry breaking 
by the diquark condensate (not by the usual chiral condensate). 
It is shown in \cite{Kanazawa:2009en} that the partition function of the RMT 
actually coincides with that of the finite-volume effective theory of QCD at large $\mu$.

\subsection{Orbifold projections in the chiral random matrix theories}
\label{sec:RMT_equivalence}
In this section, we construct the orbifold projections in the 
chiral random matrix theories (RMTs) between $\beta=4$, $\beta=2$, and $\beta=1$.
Thereby a class of observables in the RMTs with the different Dyson indices
are found to be identical to each other.
In the following, we will concentrate on the RMT at finite $\mu$ and $T=0$ 
introduced in Sec.~\ref{sec:lowRMT}, which can be easily generalized to nonzero $T$.
For simplicity, we set $\nu=0$ and consider degenerate quark masses $m_f=m$.
The generalizations to the high-density RMTs in Sec.~\ref{sec:highRMT} 
and to the RMTs in the two-matrix representation 
\cite{Osborn:2004rf, Akemann:2005fd} are straightforward.

\begin{figure}[t]
\begin{center}
\includegraphics[width=11cm]{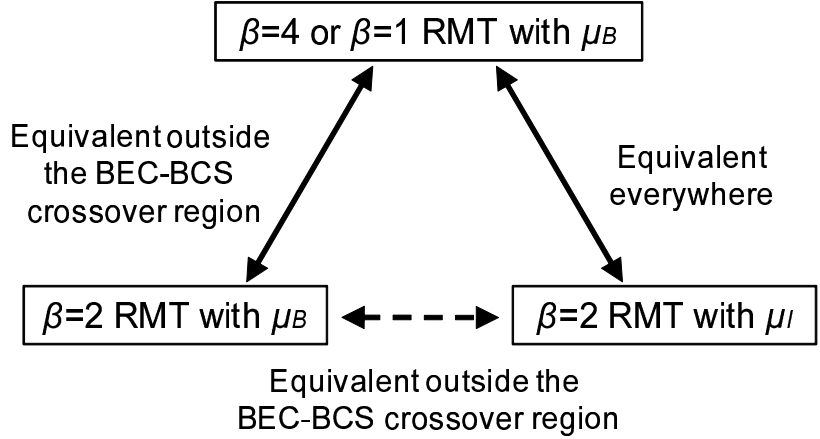}
\end{center}
\vspace{-0.5cm}
\caption{Relations between $\beta=2$ RMT at finite $\mu_B$ or $\mu_I$
and $\beta=4$ and $\beta=1$ RMTs at finite $\mu_B$. 
$\beta=2$ RMT at small and large $\mu_I$ can be obtained from $\beta=4$ and 
$\beta=1$ RMTs at small and large $\mu_B$ through orbifold projections.
$\beta=2$ RMT at small $\mu_B$ can also be obtained from $\beta=4$ and 
$\beta=1$ RMTs at small $\mu_B$ while $\beta=2$ RMT at large $\mu_B$ 
(inside the BEC-BCS crossover region of the parent RMTs) cannot be.}
\label{fig:RMT}
\end{figure}

The construction of the orbifold projections is as follows: 
we start with the $\beta=4$ or $\beta=1$ RMT at finite $\mu_B$
with the size of the matrix $\Phi$ being $2N$, 
and define the orbifold projection to the $\beta=2$ RMT
at finite $\mu_B$ or $\mu_I$ with the size of the matrix $N$.
The relations between these RMTs via orbifold projections are summarized 
in Fig.~\ref{fig:RMT}.\footnote{Although not shown explicitly in this paper, one 
can also construct the orbifold projection from the $\beta=2$ RMT at finite $\mu_I$
with the size $N$ to the $\beta=1$ RMT at finite $\mu_B$ with the size $N$.}
Each orbifold projection reduces the independent degrees of freedom 
of the theory to half.

\subsubsection{Orbifold projection from $\beta=4$ to $\beta=2$}
\label{sec:4-2}
The partition function of the $\beta=4$ RMT is given by 
\begin{eqnarray}
Z=\int d\Phi d\Psi \ e^{-S}, \qquad S=S_{B}+S_{F},
\end{eqnarray}
where 
\begin{eqnarray}
\label{eq:RMT-action}
S_B
=\frac{N\beta}{2}G^2 \tr \Phi^\dagger \Phi, \qquad
S_F
=
\sum_{f=1}^{N_f}
\bar{\Psi}_f {\cal D}\Psi_f, 
\end{eqnarray}
and  
\begin{eqnarray}
{\cal D}
=
\left(
\begin{array}{cc}
m \textbf{1}_{2N}& \Phi+\mu\textbf{1}_{2N} \\
-\Phi^\dagger +\mu\textbf{1}_{2N} & m \textbf{1}_{2N}
\end{array}
\right).   
\end{eqnarray}
Here $\Psi_f$ are complex Grassmann $4N$-component vectors and
$\Phi$ is a $2N\times 2N$ quaternion real matrix of the form:
\begin{eqnarray}
\label{eq:quaternion}
\Phi
\equiv \sum_{\mu=0}^3 a^{\mu} i \sigma_{\mu}=
\left(
\begin{array}{cc}
a^0 + i a^3 &  a^2 + i a^1 \\
-a^2 + i a^1 & a^0 - i a^3
\end{array}
\right), 
\end{eqnarray}
where $a^{\mu}$ are $N\times N$ real matrices 
and $\sigma_{\mu}=(-i, \sigma_k)$ with Pauli matrices $\sigma_k$.

For the bosonic matrix $\Phi$, we impose the projection condition 
\begin{eqnarray}
J \Phi J^{-1} = \Phi, \qquad
J \equiv
\left(
\begin{array}{cc}
& -\textbf{1}_{N}   \\
\textbf{1}_{N}  & 
\end{array}
\right).   
\end{eqnarray}
Then we obtain 
\begin{eqnarray}
\Phi^{\rm proj}
=
\left(
\begin{array}{cc}
a^0  & a^2\\
-a^2 & a^0 
\end{array}
\right), 
\end{eqnarray}
which is equivalent to two copies of a $N\times N$ complex matrix
after a unitary transformation
\begin{eqnarray}
P \Phi^{\rm proj} P^{-1} =
  \left(
\begin{array}{cc}
\phi^\ast& 0\\
0 & \phi
\end{array} 
\right) 
\equiv \Phi_{\beta=2}, \qquad
P \equiv \frac{1}{\sqrt{2}}\left(
\begin{array}{cc}
\textbf{1}_{N} & i \textbf{1}_{N} \\
\textbf{1}_{N} & -i \textbf{1}_{N}
\end{array} 
\right),
\end{eqnarray}
where $\phi = a^0+ia^2$.  
The bosonic part of the action is mapped to the one for the 
$\beta=2$ RMT. 
Note that the factor 2 in the recipe (\ref{eq:recipe}) is
reflected in the difference of normalization between $\beta=4$ and $\beta=2$
in (\ref{eq:RMT-action}) if the trace for $\beta=4$ is understood 
as the so-called ``QTr" which is one-half the usual trace.

In order to introduce a projection for the fermions, we write $\Psi$ by using two
$2N$-component fermions $\psi_R$ and $\psi_L$ as
\begin{eqnarray}
\Psi=
\left(
\begin{array}{c}
\psi_R \\
\psi_L
\end{array}
\right). 
\end{eqnarray}
Here $\psi_R$ and $\psi_L$ are further decomposed into two $N$-component fermions
\begin{eqnarray}
\psi_R=
\left(
\begin{array}{c}
\psi_R^1\\
\psi_R^2
\end{array}
\right), \qquad
\psi_L=
\left(
\begin{array}{c}
\psi_L^1\\
\psi_L^2
\end{array}
\right),
\end{eqnarray}
where the flavor index is suppressed for simplicity.
The projection to the $\beta=2$ RMT at finite $\mu_B$ is given by 
\begin{eqnarray}
\psi_R = \omega J \psi_R,
\qquad
\psi_L = \omega J \psi_L,  
\end{eqnarray}
where $\omega=e^{i\pi/2}$ as defined in Sec.~\ref{sec:projection}. 
Performing the unitary transformation, we have 
\begin{eqnarray}
P\psi_R^{\rm proj}
=
\left(
\begin{array}{c}
0\\
\psi_R^{-}
\end{array}
\right),  
\qquad
P\psi_L^{\rm proj}
=
\left(
\begin{array}{c}
0\\
\psi_L^{-}
\end{array}
\right),  
\end{eqnarray}
where $\psi_{R}^{\pm}=(\psi_R^1 \pm i\psi_R^2)/\sqrt{2}$ and 
$\psi_L^{\pm}=(\psi_L^1 \pm i\psi_L^2)/\sqrt{2}$. 
The fermionic part of the action reads
\begin{eqnarray} (0,\bar{\psi}_L^-,0,\bar{\psi}_R^-)
\left(
\begin{array}{cccc}
m \textbf{1}_{N} & 0 & \phi^\ast + \mu\textbf{1}_{N} & 0 \\
0 & m \textbf{1}_{N}& 0 &  \phi + \mu\textbf{1}_{N} \\
-\phi^T + \mu\textbf{1}_{N} & 0 & m \textbf{1}_{N} & 0 \\
0 & -\phi^\dagger +  \mu\textbf{1}_{N} & 0 & m \textbf{1}_{N}
\end{array}
\right)
\left(
\begin{array}{c}
0\\
\psi_R^-\\
0\\
\psi_L^-
\end{array} 
\right)
=2\bar{\Psi}_{\beta=2}{\cal D}(\mu)_{\beta=2}\Psi_{\beta=2},
\nonumber \\
\end{eqnarray}
where
\begin{eqnarray}
{\cal D}(\mu)_{\beta=2}
=
\left(
\begin{array}{cc}
m \textbf{1}_{N}& \phi + \mu\textbf{1}_{N} \\
-\phi^\dagger +\mu\textbf{1}_{N} & m \textbf{1}_{N}
\end{array}
\right), \qquad
\Psi_{\beta=2}=\frac{1}{\sqrt{2}}
\left(
\begin{array}{c}
\psi_R^-\\
\psi_L^-
\end{array}
\right).
\end{eqnarray}
This is the $\beta=2$ RMT at finite quark chemical potential $\mu$.

In order to obtain the $\beta=2$ RMT at finite $\mu_I$, 
we impose the projection condition:
\beq
J\psi_R J_i^{-1}=\psi_R, \qquad J\psi_L J_i^{-1}=\psi_L.
\eeq 
If we define 
$\varphi_{R}^{\pm}=(\psi_{Rf}^{\pm} \mp i \psi_{Rg}^{\pm})/\sqrt{2}$ and
$\xi_{R}^{\pm}=(\psi_{Rf}^{\pm} \pm i \psi_{Rg}^{\pm})/\sqrt{2}$
(and similarly for $\varphi_L^{\pm}$ and $\xi_L^{\pm}$)
with $f$ and $g$ being the flavor indices,
the fermions $\varphi_{R,L}^{\pm}$ survive but $\xi_{R,L}^{\pm}$
disappear after the projection.
The fermionic part of the action reads
\begin{eqnarray}
& &
(\bar{\varphi}_L^+,\bar{\varphi}_L^-,\bar{\varphi}_R^+,\bar{\varphi}_R^-)
\left(
\begin{array}{cccc}
m \textbf{1}_{N} & 0 & \phi^\ast+\mu\textbf{1}_{N} & 0 \\
0 & m \textbf{1}_{N}& 0 &  \phi+\mu\textbf{1}_{N} \\
-\phi^T + \mu\textbf{1}_{N} & 0 & m \textbf{1}_{N} & 0 \\
0 & -\phi^\dagger +  \mu\textbf{1}_{N} & 0 & m \textbf{1}_{N}
\end{array}
\right)
\left(
\begin{array}{c}
\varphi_R^+\\
\varphi_R^-\\
\varphi_L^+\\
\varphi_L^-
\end{array} 
\right)
\nonumber\\
& & 
=
2\left[\bar \Psi_{\beta=2}^{+}
{\cal D}(-\mu)_{\beta=2}
\Psi_{\beta=2}^{+}
+ \bar \Psi_{\beta=2}^{-}
{\cal D}(\mu)_{\beta=2}
\Psi_{\beta=2}^{-}\right],
\end{eqnarray}
where
\begin{eqnarray}
\Psi_{\beta=2}^{+}=\frac{1}{\sqrt{2}}
\left(
\begin{array}{c}
(\varphi_R^+)^C \\
(\varphi_L^+)^C
\end{array} 
\right), \qquad
\Psi_{\beta=2}^{-}=\frac{1}{\sqrt{2}}
\left(
\begin{array}{c} 
\varphi_R^-\\
\varphi_L^-
\end{array} 
\right).
\end{eqnarray}
Because $\Psi^-$ and $\Psi^+$ have the quark chemical potential $+ \mu$ 
and $- \mu$ respectively,
it is the $\beta=2$ RMT at finite isospin chemical potential $\mu_I=2\mu$.

\subsubsection{Orbifold projection from $\beta=1$ to $\beta=2$}
The $\beta=2$ RMT can also be obtained from the $\beta=1$ RMT.
We start with the action of the $\beta=1$ RMT given by
(\ref{eq:RMT-action}), but the Dirac operator is now
\begin{eqnarray}
{\cal D}
=
\left(
\begin{array}{cc}
m \textbf{1}_{2N}& \Phi+\mu\textbf{1}_{2N} \\
-\Phi^T +\mu\textbf{1}_{2N} & m \textbf{1}_{2N}
\end{array}
\right),   
\end{eqnarray}
where $\Phi$ is a $2N\times 2N$ real matrix. 
$\Phi$ can be parametrized as
\begin{eqnarray}
\Phi =
\left(
\begin{array}{cc}
a^0 +  a^3 &  a^2 +  a^1 \\
-a^2 +  a^1 & a^0 -  a^3
\end{array}
\right), 
\end{eqnarray}
where $a^{\mu}$ are $N\times N$ real matrices. Note that the only change 
in this expression compared with (\ref{eq:quaternion}) is that the 
factors $i$ in front of $a^0$ and $a^3$ are absent.
Then one can easily find that the same projection conditions for $\Phi$
and $\Psi$ as Sec.~\ref{sec:4-2} give the $\beta=2$ RMT 
at finite $\mu$ or finite $\mu_I$.

\subsection{Solving the chiral random matrix theories}
\label{sec:solveRMT}
The orbifold equivalence in the RMT predicts that the $\beta=4$ and $\beta=1$ RMTs
at finite $\mu_B$ and $\beta=2$ RMT at finite $\mu_I$ are equivalent to each other 
in the neutral sector. Moreover, outside the superfluid phase,
the above three theories must also be equivalent to the $\beta=2$ RMT at finite $\mu_B$.
In this section, we will verify these statements at the nonperturbative level by computing 
the effective potential of each RMT. 

For the $\beta=2$ and $\beta=1$ RMTs, the effective potentials are computed in 
\cite{Klein:2003fy, Klein:2004hv}. The equivalence of the effective potential
of the $\beta=2$ RMT at finite $\mu_B$ (at $\mu_I=0$) and that of the $\beta=2$ RMT at finite $\mu_I$ 
(at $\mu_B=0$) is pointed out outside the pion condensed phase.
Here we show that the equivalence holds between a larger class of RMTs
as a natural consequence of the orbifold projections constructed in the previous subsection.

Let us sketch the derivation of the effective potential of the RMT.
First, introduce the Grassmann vectors (fermions) $\psi$ to write the determinant 
into an exponential form.
Second, perform the Gaussian integration over the matrix element $\Phi$, 
which leads to the four-fermion term in the exponent.
Third, introduce the bosonic auxiliary field $A$ to make it the fermion bilinears 
(this procedure is called the Hubbard-Stratonovich transformation).
Fourth, perform the Gaussian integration over $\Psi$. Finally, the effective potential
is given as the saddle point of the integrand in the $N \rightarrow \infty$ limit
(the thermodynamic limit).\footnote{Our arguments depend on the ansatz 
of the saddle point of $A$ (defined below) 
at finite $T$ and $\mu$ similarly to \cite{Klein:2003fy, Klein:2004hv}.}

In the following, we consider the $N_f=2$ RMT in (\ref{eq:RMT})
with the quark mass $m_f$ and the chemical potential 
$\mu_f$ for each flavor, $f=1,2$.
The baryon and isospin chemical potentials are defined as
\beq
\bar \mu_B &\equiv & \frac{\mu_B}{N_c} = \frac{1}{2}(\mu_1 + \mu_2),
\\
\bar \mu_I &\equiv & \frac{\mu_I}{2} = \frac{1}{2}(\mu_1 - \mu_2).
\eeq
We denote the chiral condensate as $\sigma_f$, pion condensate as $\rho$, 
diquark condensate as $\Delta$, and their sources as $m_f$, $\lambda$, 
and $J$, respectively.

\subsubsection{Effective potential of $\beta=4$}
We first consider the $\beta=4$ RMT with degenerate quark mass $m_f=m$ 
at finite baryon chemical potential $\mu_f = \bar \mu_B$ 
(and hence $\sigma_f=\sigma$).
We will focus on $T=0$ and introduce $T$ later changing as 
$\bar \mu_B \rightarrow \bar \mu_B + iT$ and $\bar \mu_B \rightarrow \bar \mu_B - iT$ 
for each half of the fermion determinant following \cite{Vanderheyden:2005ux}.

Remembering the definition of the quaternion real matrix (\ref{eq:quaternion}),
the partition function can be rewritten as
\beq
\label{eq:Z1}
Z &=& \int da^{\mu} d \psi^* d \psi \ 
\exp \left[- 2N G^2 \tr (a_{\mu} a_{\mu}^T) 
+ \psi^{f*}_{Ri} (a_{ij}^{\mu} i \sigma_{\mu} + \bar \mu_B {\bf 1}_{ij})\psi^{f}_{Lj} \right.
\\
& & \ \qquad \qquad \qquad \qquad
- \psi^{g}_{Ri} (a_{ij}^{\mu} i \sigma_{\mu}^{*} + \bar \mu_B {\bf 1}_{ij})\psi^{g*}_{Lj} 
+M^{\dag}_{fg} \psi^{f*}_{Ri} \psi^g_{Ri} + 
M_{fg} \psi^{f*}_{Lj} \psi^g_{Lj} \Bigr],
\nonumber 
\eeq
where $\psi_{R,L}$ are the Grassmann $2N$-component vectors
and $f,g$ are flavor indices and $i,j$ run over $1,2,\cdots 2N$.
Integrating out $a_{\mu}$ and using the Fierz identity,
\beq
(\sigma_{\mu})^{ab} (\sigma_{\mu}^{*})^{cd}
=2\delta_{ac}\delta_{db},
\eeq
the partition function reduces to
\beq
Z &=& \int d \psi^* d \psi \ 
\exp \left[\frac{1}{2NG^2} (\psi^{f *}_{Ri}\psi^{g }_{Ri}) 
(\psi^{g *}_{Lj}\psi^{f}_{Lj})
+ \bar \mu_B 
(\psi^{f *}_{Ri} \psi^{f}_{Li} + \psi^{g *}_{Lj} \psi^{g}_{Rj}) \right.
\nonumber \\
& & \qquad \qquad \qquad \qquad +M^{\dag}_{fg} \psi^{f *}_{Ri} \psi^{g}_{Ri} + 
M_{fg} \psi^{f *}_{Lj} \psi^{g}_{Lj} \biggr].
\eeq
Performing the Hubbard-Stratonovich transformation by introducing the auxiliary 
real and symmetric $N_f \times N_f$ matrices $K_{fg}$ and $L_{fg}$, one has
\beq
Z &=& \int d K d L d \psi^* d \psi
\exp \left[- 8 NG^2 \tr (K^2 + L^2) \right.
+ \bar \mu_B (\psi^{f *}_{Ri} \psi^{f}_{Li} + \psi^{g *}_{Lj} \psi^{g}_{Rj})
\nonumber \\
& & \qquad \qquad \qquad \qquad \quad
+ 2 \psi^{f *}_{Ri}\psi^{g}_{Ri}(K+iL)_{fg}
+ 2 \psi^{g*}_{Lj} \psi^{f}_{Lj}(K-iL)_{fg} 
\nonumber \\
& & \qquad \qquad \qquad \qquad \quad
+M^{\dag}_{fg} \psi^{f *}_{Ri} \psi^{g}_{Ri} + 
M_{fg} \psi^{f *}_{Lj} \psi^{g}_{Lj}\Bigr].
\eeq
Integrating over the fermionic variables $\psi$ and 
$\psi^*$ leads to the expression:
\beq
Z=\int dA \exp[-N\Omega_{\beta=4}(A,A^{\dag})],
\eeq
where $\Omega_{\beta=4}$ is an effective potential given by
\beq
\Omega_{\beta=4}= 8 G^2 \tr(A A^{\dag}) - 2 \log \det Q.
\eeq
Here $A=K-iL$ is the complex and symmetric $N_f \times N_f$ matrix 
and
\beq
Q=\left(
\begin{array}{cccc}
2A^{\dag} + M^{\dag} & \bar \mu_B \delta_{fg} \\
\bar \mu_B \delta_{fg} & 2A + M
\end{array}
\right).
\eeq
We set the source term $M$ and make the ansatz for $A$ as follows,
\beq
M = \left( {\begin{array}{*{20}c}
    m  & {i J}  \\
   {i J} & m   \\
\end{array}} \right), \qquad
A = \left( {\begin{array}{*{20}c}
   \sigma  & {i\Delta }  \\
   {i\Delta } & \sigma   \\
\end{array}} \right).
\eeq
Shifting $\sigma$ and $\Delta$ such that $m$ and $J$ dependences are 
absorbed into the quadratic term and adding the $T$-dependence,
one finally arrives at the effective potential of $\beta=4$ RMT:
\beq
\label{eq:potential4}
\Omega_{\beta=4} = 16G^2\left[ \left(\sigma- \frac{m}{2} \right)^2 
+ \left(\Delta- \frac{J}{2} \right)^2 \right] 
- 2 \sum_{\pm} \ln [4\sigma^2 + 4\Delta^2 - (\bar \mu_B \pm iT)^2].
\eeq
The chiral condensate and the diquark condensate are expressed using 
$\sigma$ and $\Delta$ as
\beq
\langle \bar u u \rangle_{\beta=4} 
&=& \left. \frac{1}{4N}\partial_{m}\ln Z_{\beta=4} \right|_{m=0} = - 4G^2 \sigma_{\beta=4},
\\
\langle u^T C \gamma_5 u \rangle_{\beta=4} 
&=& \left. \frac{1}{4N}\partial_{J}\ln Z_{\beta=4} \right|_{J=0} = - 4G^2 \Delta_{\beta=4}.
\eeq

\subsubsection{Effective potential of $\beta=2$}
Similarly to the case with $\beta=4$, one can obtain
the effective potential of $\beta=2$ RMT. This was previously computed
in \cite{Klein:2003fy} and the result reads
\beq
\label{eq:potential2}
\Omega_{\beta=2} &=& G^2[(\sigma_1-m_1)^2+(\sigma_2-m_2)^2 + 2(\rho-\lambda)^2]
\nonumber \\
	& & -\frac{1}{2}\sum_{\pm}
\ln [(\sigma_1 + \mu_1 \pm iT)(\sigma_2 - \mu_2 \mp iT) + \rho^2]
[(\sigma_1 - \mu_1 \mp iT)(\sigma_2 + \mu_2 \pm iT) + \rho^2].
\nonumber \\
\eeq
The chiral condensate and pion condensate are calculated as
\beq
\langle \bar u u \rangle_{\beta=2} 
&=& \left. \frac{1}{2N}\partial_{m_1}\ln Z_{\beta=2} \right|_{m_1=0} = - G^2 \sigma_{\beta=2},
\\
\langle \bar d \gamma^5 u \rangle_{\beta=2} 
&=& \left. \frac{1}{4N}\partial_{\lambda}\ln Z_{\beta=2} \right|_{\lambda=0} = - G^2 \rho_{\beta=2}.
\eeq
Note that, as long as $\rho=0$ (i.e., outside the pion condensed phase), 
the potential (\ref{eq:potential2}) is a function of $\mu_1^2=(\bar \mu_B + \bar \mu_I)^2$ 
and $\mu_2^2 = (\bar \mu_B - \bar \mu_I)^2$. This property leads to the relation:
\beq
\label{eq:equivalence2}
\Omega_{\beta=2}(\bar \mu_B)|_{\bar \mu_I=0}
=\Omega_{\beta=2}(\bar \mu_I)|_{\bar \mu_B=0} \ \ {\rm for} \ \ {\rho=0}.
\eeq
Here $\rho=0$ is the condition that the projection symmetry,
which is used for the orbifolding in Sec.~\ref{sec:RMT_equivalence}, 
is not broken spontaneously.
From (\ref{eq:equivalence2}), the magnitude of the chiral condensate $\sigma$ and 
the critical temperature of chiral phase transition $T^{\sigma}$ in each theory coincide,
\beq
\sigma_{\beta=2}(\bar \mu_B)|_{\bar \mu_I=0}
&=&\sigma_{\beta=2}(\bar \mu_I)|_{\bar \mu_B=0} \ \ {\rm for} \ \ {\rho=0},
\\
T^{\sigma}_{\beta=2}(\bar \mu_B)|_{\bar \mu_I=0}
&=&T^{\sigma}_{\beta=2}(\bar \mu_I)|_{\bar \mu_B=0} \ \ {\rm for} \ \ {\rho=0},
\eeq
as a consequence of the orbifold equivalence.
Especially, this shows that the phase-quenched approximation for $\sigma_{\beta=2}(\bar \mu_B)$
and $T^{\sigma}_{\beta=2}(\bar \mu_B)$ works outside the pion condensed phase, 
as mentioned in \cite{Klein:2003fy}. 

It should be remarked that, 
even though the effective potentials are identical in (\ref{eq:equivalence2}) for $\rho=0$,
the partition functions themselves are not generally the same. This is because the 
preexponential factor also contributes to the partition function, 
which is not taken into account in computing the effective potential.\footnote{
For $\rho \neq 0$, the partition functions are not identical even in the leading exponential behavior.}
Therefore, the sign problem measured as the phase of the partition function
can be severe inside as well as outside the pion condensed phase \cite{Han:2008xj}. 
The result here shows that the phase-quenched approximation is exact for the observables 
above independently of the severity of the sign problem, as long as $\rho=0$.

\subsubsection{Effective potential of $\beta=1$}
The effective potential of $\beta=1$ RMT is computed in \cite{Klein:2004hv} as
\beq
\label{eq:potential1}
\Omega_{\beta=1} &=& G^2[(\sigma_1-m_1)^2+(\sigma_2-m_2)^2 + 2(\rho-\lambda)^2 + 2(\Delta-J)^2]
\nonumber \\
	& & -\frac{1}{4}\sum_{\pm}
\ln \{[(\sigma_1 + \mu_1 \pm iT)(\sigma_2 - \mu_2 \mp iT) + \rho^2 + \Delta^2 ]
\nonumber \\
& & \qquad \qquad \times[(\sigma_1 - \mu_1 \pm iT)(\sigma_2 + \mu_2 \mp iT) + \rho^2 + \Delta^2] + 4\Delta^2 \mu_1 \mu_2 \}
\nonumber \\
& & \qquad \qquad \times \{[(\sigma_1 - \mu_1 \mp iT)(\sigma_2 + \mu_2 \pm iT) + \rho^2 + \Delta^2 ]
\nonumber \\
& & \qquad \qquad \times (\sigma_1 + \mu_1 \mp iT)(\sigma_2 - \mu_2 \pm iT) + \rho^2 + \Delta^2] + 4\Delta^2 \mu_1 \mu_2 \}.
\eeq
The chiral condensate, pion condensate, and diquark condensate read
\beq
\langle \bar u u \rangle_{\beta=1} 
&=& \left. \frac{1}{2N}\partial_{m_1}\ln Z_{\beta=1} \right|_{m_1=0} = - G^2 \sigma_{\beta=1},
\\
\langle \bar d \gamma^5 u \rangle_{\beta=1} 
&=& \left. \frac{1}{4N}\partial_{\lambda}\ln Z_{\beta=1} \right|_{\lambda=0} = - G^2 \rho_{\beta=1}.
\\
\langle d^T C \gamma_5 u \rangle_{\beta=1} 
&=& \left. \frac{1}{4N}\partial_{J}\ln Z_{\beta=1} \right|_{J=0} = - G^2 \Delta_{\beta=1}.
\eeq
The potential (\ref{eq:potential1}) has the symmetry
\beq
\Omega_{\beta=1}(\Delta, \rho, \mu_1, \mu_2)
=\Omega_{\beta=1}(\rho, -\Delta, \mu_1, -\mu_2),
\eeq
due to the $\bar \mu_B \leftrightarrow \bar \mu_I$ symmetry for $\beta=1$.
Note that this symmetry has nothing to do with the orbifold equivalence.

\subsubsection{Nonperturbative orbifold equivalence between $\beta=4$, $\beta=2$, and $\beta=1$}
Comparing (\ref{eq:potential4}), (\ref{eq:potential2}), and (\ref{eq:potential1}),
and using the $\bar \mu_B \leftrightarrow \bar \mu_I$ symmetry for $\beta=1$ RMT, 
one finds the relation 
(note that $\Delta=0$ at $\bar \mu_B=0$ and $\rho=0$ at $\bar \mu_I=0$):
\begin{gather}
\label{eq:EP}
\Omega_{\beta=4}(\bar \mu_B)|_{\bar \mu_I=0}
= 2\Omega_{\beta=2}(\bar \mu_I)|_{\bar \mu_B=0}
= 2\Omega_{\beta=1}(\bar \mu_B)|_{\bar \mu_I=0},
\\
\label{eq:argument}
{\rm for} \ \ 2\sigma_{\beta=4}=\sigma_{\beta=2}=\sigma_{\beta=1}, \ \ 
2\Delta_{\beta=4}=\rho_{\beta=2}=\Delta_{\beta=1}.
\end{gather}
Unlike the relation (\ref{eq:equivalence2}), this
is valid not only for $\rho=0$ (or $\Delta=0$) 
but also for $\rho \neq 0$ (or $\Delta \neq 0$),
as is consistent with our claim in Sec.~\ref{sec:condition}. 
The difference of the factor 2 in (\ref{eq:EP})
originates from the recipe (\ref{eq:recipe}), or
the fact that the $\beta=4$ RMT with the size of the matrix $\Phi$ being $2N$
includes 2 copies of the $\beta=2$ or $\beta=1$ RMT with the size $N$.
The factors 2 in (\ref{eq:argument})
originate from the identifications of the chiral condensate and
diquark (or pion) condensate according to the recipe (\ref{eq:recipe}) again, 
e.g., $\langle \bar u u \rangle_{\beta=4}=2\langle \bar u u \rangle_{\beta=2}
=2\langle \bar u u \rangle_{\beta=1}$.
The relation (\ref{eq:EP}) leads to the equivalence of the magnitudes of the order parameters
(up to the factor 2) and the critical temperatures:
\beq
2\sigma_{\beta=4}(\bar \mu_B)|_{\bar \mu_I=0}
=\sigma_{\beta=2}(\bar \mu_I)|_{\bar \mu_B=0}
=\sigma_{\beta=1}(\bar \mu_B)|_{\bar \mu_I=0},
\\
2\Delta_{\beta=4}(\bar \mu_B)|_{\bar \mu_I=0}
=\rho_{\beta=2}(\bar \mu_I)|_{\bar \mu_B=0}
=\Delta_{\beta=1}(\bar \mu_B)|_{\bar \mu_I=0},
\\
T^{\sigma}_{\beta=4}(\bar \mu_B)|_{\bar \mu_I=0}
=T^{\sigma}_{\beta=2}(\bar \mu_I)|_{\bar \mu_B=0}
=T^{\sigma}_{\beta=1}(\bar \mu_B)|_{\bar \mu_I=0},
\\
T^{\Delta}_{\beta=4}(\bar \mu_B)|_{\bar \mu_I=0}
=T^{\rho}_{\beta=2}(\bar \mu_I)|_{\bar \mu_B=0}
=T^{\Delta}_{\beta=1}(\bar \mu_B)|_{\bar \mu_I=0},
\eeq
as a consequence of the orbifold equivalence.
We note that, the equivalence of the neutral order parameters and the
critical temperatures should be satisfied in the original QCD and QCD-like theories
as we claimed in Sec.~\ref{sec:QCD}, while 
the effective potentials will not necessarily coincide in QCD as (\ref{eq:EP}).
In the case of the RMT, the effective potential is a 
function of only the neutral order parameters and
all the moments are identical; 
as a result, the effective potentials must be identical. 
In QCD and QCD-like theories, the effective potentials are 
functions of not only the neutral order parameters but also 
non-neutral ones, so the effective potentials will not be identical generally.

\subsection{Brief summary}
We have applied the idea of the orbifold equivalence to the chiral random
matrix theories (RMTs) and constructed the orbifold projections between the RMTs
with different Dyson indices $\beta$. The equivalence of the order parameters,
both the chiral condensate and diquark (or pion) condensate,
has been demonstrated by computing the effective potentials of RMTs.

From the viewpoint of the orbifold projection, the construction of 
whole class of the RMTs at finite $\mu$ can be understood in a unified way (see Fig.~\ref{fig:RMT}).
However, the $\beta=2$ RMT at large $\mu_B$ is the only exception among these RMTs
which cannot be obtained through the orbifold projection from the parent RMT
because of the spontaneous breaking of the projection symmetry.
This may be the fundamental reason why an RMT at large $\mu_B$ 
which reproduces the partition function of the color-flavor locked phase 
in the $\epsilon$-regime \cite{Yamamoto:2009ey} has not been constructed yet.

\section{Conclusion and discussions}
\label{sec:conclusion}
In this paper, we have discussed the universality of the phase diagrams of 
QCD and QCD-like theories in the large-$N_c$ limit via the orbifold equivalence. 
The whole phase diagrams described by the chiral condensate and the superfluid gap
are identical between $\SU(N_c)$ QCD at finite isospin chemical potential $\mu_I$ and 
$\SO(2N_c)$ and $\Sp(2N_c)$ gauge theories 
at finite baryon chemical potential $\mu_B$.
The phase diagrams of these theories outside the BEC-BCS crossover regions
are also identical to that of $\SU(N_c)$ QCD at finite $\mu_B$.
Especially, the chiral condensate and its critical temperature 
in QCD at finite $\mu_B$ should be exactly described by those of sign-free QCD 
at finite $\mu_I$: the phase-quenched approximation for these quantities 
is exact in the large-$N_c$ limit outside the BEC-BCS crossover region.\footnote{One might 
suspect that, for $\mu<m_{\pi}/2$ and small but finite $T$, 
the system is a gas of baryons in QCD at finite $\mu_B$, which is completely different 
from a gas of pions in QCD at finite $\mu_I$, and the equivalence would not hold.
However, this difference is irrelevant to the orbifold equivalence of 
the neutral operators in the large-$N_c$ limit. 
For example, consider the chiral condensate at small $T$. In QCD at finite $\mu_B$,
the thermal excitation of heavy baryons with the mass $\sim N_c^1$ 
is suppressed at small $T = {\cal O} (N_c^0)$ so that the chiral condensate remains 
unchanged from the value in the QCD vacuum.
On the other hand, in QCD at finite $\mu_I$, the thermal excitation of pions 
with the mass $\sim N_c^0$ cannot change the chiral condensate $\sim N_c^1$.
Therefore, the magnitude of the chiral condensate should be the same in both theories 
in this region, and the equivalence is satisfied rather trivially.
The prediction of the orbifold equivalence is nontrivial at larger $T$ near 
the critical temperature $T_c(\mu)$ of the chiral phase transition;
if it is larger than the critical temperature of the deconfinement phase transition,
it generally depends on $\mu$, for which the orbifold equivalence 
still predicts the exactly the same $T_c(\mu)$ in both theories.}
We have also checked that the equivalence is well satisfied for $N_c=3$ 
at asymptotically high densities using the controlled weak-coupling calculations.
This leads us to expect that the phase-quenched approximation for the
chiral condensate also works well even for $N_c=3$.

Our results provide a way to evade the sign problem in the lattice QCD 
simulation at finite $\mu_B$, especially in the region relevant to the 
physics of the chiral phase transition at high temperature.
The putative QCD critical point may be investigated by studying the sign-free
QCD at finite $\mu_I$ and $\SO(2N_c)$ and $\Sp(2N_c)$ gauge theories at finite $\mu_B$ 
outside the BEC-BCS crossover regions.
The lattice QCD simulations in QCD at finite $\mu_I$ were already performed 
in \cite{deForcrand:2007uz, Kogut:2007mz} which seem consistent with 
the results at finite $\mu_B$ \cite{deForcrand:2006pv, deForcrand:2008vr}
though they may not be conclusive. 
Further investigations in this direction would be desirable.

Other interesting phenomena in QCD at large $\mu_B$ and at low $T$, 
such as the color superconductivity \cite{Alford:2007xm} 
and the quarkyonic phase \cite{McLerran:2007qj, Kojo:2009ha},
are unfortunately inside the BEC-BCS crossover region of the parent 
$\SO$ and $\Sp$ gauge theories where the orbifold equivalence breaks down.
One may add appropriate deformation to $\SO(2N_c)$ \cite{Cherman:2010jj, Cherman:2011mh} 
and $\Sp(2N_c)$ gauge theories to prevent the BEC of diquark pairing at small $\mu$, 
which allows us to study the properties of QCD beyond $\mu=m_{\pi}/2$ 
using the lattice technique.
One should note, however, that it is this BEC-BCS crossover region 
inside which the QCD phase diagram at finite $\mu_B$ crucially depends on $N_c$.
For example, the color superconductivity is no longer energetically favorable
and is replaced by the inhomogeneous chiral density wave in the 't Hooft limit 
\cite{Deryagin:1992rw, Shuster:1999tn}.
This is in contrast to QCD at finite $\mu_B$ outside the BEC-BCS crossover region, 
QCD at finite $\mu_I$, and $\SO(2N_c)$ and $\Sp(2N_c)$ gauge theories
at finite $\mu_B$, where the phase structures are not affected by $N_c$ 
qualitatively.
There might be some connection between the region where the physics dramatically changes 
depending on $N_c$ and the region where the orbifold equivalence breaks down.

As we have revealed in this paper, the idea of the orbifold equivalence
is useful to discuss the universal properties of different quantum field theories.
One should be able to see the universality of phase diagrams in QCD and QCD-like 
theories within a holographic model of QCD,
the Sakai-Sugimoto model \cite{Sakai:2004cn} and 
its generalizations to $\SO(2N_c)$ and $\Sp(2N_c)$ gauge groups \cite{Imoto:2009bf}.
It would also be interesting to argue possible universal properties of other systems than QCD. 
For example, one may generalize the orbifold equivalence in 
the chiral random matrix theories to other class of random matrix theories 
(Wigner-Dyson type and Bogoliubov-de Gennes type) relevant to other systems.\\

\noindent {\bf Note added}\\
Further study \cite{Hidaka:2011jj} reveals that the QCD critical point is theoretically 
ruled out in QCD at finite $\mu_B$ outside the BEC-BCS crossover regions in the 
corresponding phase diagrams of QCD at finite $\mu_I$ and $\SO(2N_c)$ and 
$\Sp(2N_c)$ gauge theories at finite $\mu_B$, at least in the large-$N_c$ limit.

\section*{Acknowledgement} 
The authors would like to thank A.~Cherman, C.~Hoyos-Badajoz, 
A.~Karch, B.~Tiburzi, and L.~Yaffe for stimulating discussions and comments.
The works of M.~H. and N.~Y. are supported by Japan Society for the Promotion 
of Science Postdoctoral Fellowships for Research Abroad. 
\appendix

\end{document}